\documentclass[apj]{emulateapj}
\usepackage{graphicx} 
\newcommand{\ha}{H$\alpha$}
\newcommand{\hb}{H$\beta$}

\newcommand{\nii}{[N\thinspace{II}]}
\newcommand{\oii}{[O\thinspace{II}]}
\newcommand{\oiii}{[O\thinspace{III}]}
\newcommand{\sii}{[S\thinspace{II}]}

\begin{document}
\title{Deep Spectroscopy of Ultra-Strong Emission Line Galaxies\altaffilmark{1,2}}
\author{
Esther~M.~Hu,$\!$\altaffilmark{3} 
Lennox~L.~Cowie,$\!$\altaffilmark{3} 
Yuko~Kakazu,$\!$\altaffilmark{4}
Amy~J.~Barger,$\!$\altaffilmark{5,3,6} 
}

\altaffiltext{1}{Based in part on data obtained at the Subaru Telescope,
   which is operated by the National Astronomical Observatory of Japan.}
\altaffiltext{2}{Based in part on data obtained at the W. M. Keck
   Observatory, which is operated as a scientific partnership among the
   the California Institute of Technology, the University of
   California, and NASA and was made possible by the generous financial
   support of the W. M. Keck Foundation.}
\altaffiltext{3}{Institute for Astronomy, University of Hawaii,
   2680 Woodlawn Drive, Honolulu, HI 96822.}
\altaffiltext{4}{Institut d'Astrophysique, Paris, 98 bis
   Boulevard Arago, F-75014 Paris.}
\altaffiltext{5}{Department of Astronomy, University of
   Wisconsin-Madison, 475 North Charter Street, Madison, WI 53706.}
\altaffiltext{6}{Department of Physics and Astronomy,
   University of Hawaii, 2505 Correa Road, Honolulu, HI 96822.}

\shorttitle{ULTRA-STRONG EMISSION LINE GALAXIES}
\shortauthors{Hu et al.\/}

\slugcomment{Submitted to the Astrophysical Journal}

\begin{abstract}

Ultra-strong emission-line galaxies (USELs) with extremely high
equivalent widths (EW(H$\beta) \ge 30$~\AA) can be used to pick out
galaxies of extremely low metallicity in the $z=0-1$ redshift range.
Large numbers of these objects are easily detected in deep narrow band
searches and, since most have detectable \oiii$\lambda$4363, their
metallicities can be determined using the direct method. These large
samples hold out the possibility for determining whether there is a
metallicity floor for the galaxy population. In this, the second of our
papers on the topic, we describe the results of an extensive
spectroscopic follow-up of the \citet{kakazu} catalog of 542 USELs
carried out with the DEIMOS spectrograph on Keck.  We have obtained high
S/N spectra of 348 galaxies.  The two lowest metallicity galaxies in our
sample have 12+log(O/H) = 6.97$\pm0.17$ and 7.25$\pm0.03$ -- values
comparable to the lowest metallicity galaxies found to date. We
determine an empirical relation between metallicity and the R23
parameter for our sample, and we compare to this to the relationship for
low redshift galaxies.  The determined metallicity-luminosity relation
for this sample is compared with that of magnitude selected samples in
the same redshift range. The emission line selected galaxies show a
metal-luminosity relation where the metallicity decreases with
luminosity and they appear to define the lower bound of the galaxy
metallicity distribution at a given continuum luminosity. We also
compute the H$\alpha$ luminosity function of the USELs as a function of
redshift and use this to compute an upper bound on the Ly$\alpha$
emitter luminosity function over the $z=0-1$ redshift range.
\end{abstract}


\section{Introduction}
\label{secintro}

Developing a large sample of low metallicity galaxies is of considerable
interest for clues it can provide to the early stages of galaxy
formation and chemical enrichment --- such as whether forming galaxies
have a baseline metallicity that reflects the early chemical enrichment
of the intergalactic medium.  The most metal-poor systems currently
known are the low redshift blue compact emission-line galaxies such as I
Zw 18 and SBS 0335-052W with measured 12+log (O/H) of $\sim7.1-7.2$
(\citealt{sar70,thuan05,izo05}).  However, despite enormous efforts,
only a few dozen xMPGs (objects with (12+log (O/H) $< 7.65$ or $Z <
Z_\sun/12$; \citealt{knia03, izo06a}) are known
\citep[e.g.,][]{oey06,izoconf}. There are too few of these for detailed
studies of the metallicity distribution function.

In the first paper of the present series we showed that large samples of
low metallicity objects can be found by searching for ultra-strong
emission-line galaxies (USELs) using very deep narrowband surveys
\citep{kakazu}.  The narrowband method has many advantages; it is an
efficient way to develop very large samples of objects and, since it
probes to much deeper line-flux limits than objective prism or continuum
surveys, we can study populations out to near redshift $z\sim1$ where
the cosmic star formation rates peak.

In the present work we report on the detailed spectroscopic follow-up of
the catalog of 542 USEL galaxies given in \citet{kakazu}. The
\citeauthor{kakazu} sample was chosen from a set of narrowband images
obtained with the SuprimeCam mosaic CCD camera on the Subaru 8.2-m using
two $\sim$120 \AA\ (FWHM) filters centered at nominal wavelengths of
8150~\AA\ and 9140~\AA\ in regions of low sky background between the OH
bands. The total covered area is 0.5 square degrees. The catalog
consists of all objects with narrow band magnitudes less than 25 and a
narrow band excess of 0.8 magnitudes for the 8150~\AA\ filter and of 1.0
magnitudes in the 9140~\AA\ filter compared to their respective
reference continuum bands. The sample should contain all galaxies in
the fields with observed frame equivalent widths significantly greater
than 120\AA\ and 180\AA\ in the two filters, and with line fluxes above
$\sim1.5\times10^{-17}$ erg cm$^{-2}$ s$^{-1}$.

Spectroscopic redshifts have now been obtained for 299 of these galaxies
using multi-object masks with the DEIMOS spectrograph {\citep{deimos03}}
on the 10-m Keck II telescope.  The observations, flux calibration, and
equivalent width measurements are discussed in $\S2$, where the
resulting measured line ratios are also described. In $\S3$ we analyse
the metallicities and describe an empirical relation for the metallicity
as a function of the R23 parameter, which we compare with low redshift
determinations of the relation. The distribution of metallicities as a
function of the properties of the galaxies is given in $\S4$ where we
also compare the results with those in magnitude selected samples at the
same redshift.  In $\S5$ we show that the results can be used to place
upper limits on the Lyman alpha emitter (LAE) luminosity function in the
$z=0-1$ redshift range, and compare this both with the local GALEX
measurement of \citet{deh08} and with higher redshift LAE functions.  A
final summary discussion is given in $\S6$. We use a standard $H_{0}$ =
70 km s$^{-1}$ Mpc$^{-1}$, $\Omega_{m}$ = 0.3, $\Omega_{\Lambda}$ = 0.7
cosmology throughout the paper.

  \begin{figure} 
  \vspace*{-1.8in}
  \begin{center}
    \includegraphics[viewport=20 0 486 626,width=3.1in,clip,angle=0,scale=0.95]{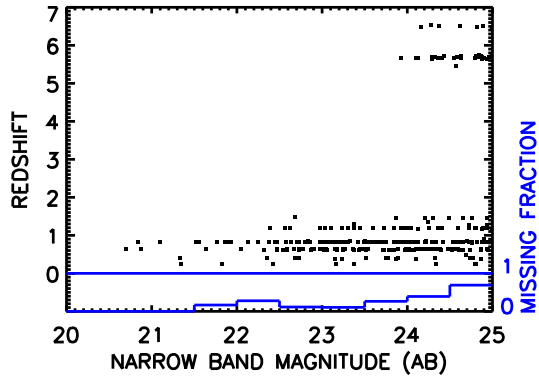}
    \caption{Redshift versus narrow band magnitude in the selection filter.
             The lower histogram (blue line) shows the fraction 
             of objects in the magnitude range which have either not 
             been observed or not been identified.
  \label{z_nmag}}
  \end{center}
  \end{figure}

\section{Spectrosopic observations}
\label{secspec}

\citep{kakazu} reported spectroscopic observations for 161 USELs from
their sample. These spectra were obtained using the Deep Extragalactic
Imaging Multi-Object Spectrograph (DEIMOS; \citealt{deimos03}) on the
Keck~II 10-m telescope in a series of runs between 2003 and 2006.  Over
the 2007 and 2008 period we have roughly doubled this spectroscopic
sample and have also significantly deepened the spectra of many of the
objects that had previously been observed.

In order to provide the widest possible wavelength coverage the
observations were primarily made with the ZD600 $\ell$/mm grating.  We
used $1''$ wide slitlets which in this configuration give a resolution
is 4.5 \AA, sufficient to distinguish the \oii $\lambda$3727 doublet
structure. This allows us to easily identify \oii $\lambda$3727 emitters
where often the \oii $\lambda$3727 doublet is the only emission feature.
The spectra cover a wavelength range of approximately 5000~\AA\ with an
average central wavelength of $7200$~\AA, though the exact wavelength
range for each spectrum depends on the slit position with respect to the
center of the mask along the dispersion direction.  The observations
were not generally taken at the parallactic angle, since this was
determined by the mask orientation, so considerable care must be taken
in measuring line fluxes, as we discuss below.  Each $\sim 1$~hr
exposure was broken into three subsets, with the objects stepped along
the slit by $1.5''$ in each direction. Some USELs were observed multiple
times, resulting in total exposure times of up to $10$ hours.  The
two-dimensional spectra were reduced following the procedure described
in \citet{cowie96} and the final one-dimensional spectra were extracted
using a profile weighting based on the strongest emission line in the
spectrum.

Our primary goal was to obtain a nearly complete set of spectra for the
200 objects in the catalog with narrow-band magnitudes brighter than 24,
since, for these brighter objects, we can obtain high quality spectra
suitable for detailed analysis. However, we also observed a large number
of the galaxies with magnitudes in the 24-25 range. We have now observed
171 of the galaxies brighter than 24. Nearly all of the bright emission
line candidates which were observed were identified (164 of the 171
objects).  However, three of the objects in the sample are stars where
the absorption line structure mimics emission in the band. The redshift
distribution of the sample and the fraction of objects which have so far
been identified is shown as a function of the narrow-band magnitude in
Figure~\ref{z_nmag}.

  \begin{figure} [t]
  \vspace*{-1.75in}
  \begin{centering}
    \includegraphics[viewport=20 0 486 626,width=3.4in,clip,angle=0,scale=0.85]{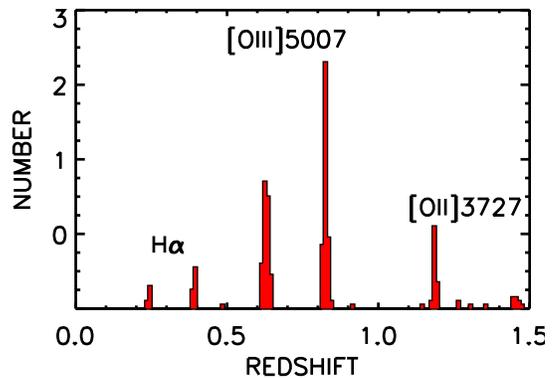}
    \caption{Redshift distribution of the spectroscopically observed
            sample. The peaks correspond to the positions at which
            the strong emission lines cross the two filters. Thus
            there are galaxies at $z=0.6$ and $z=0.8$ corresponding
            to the \oiii$\lambda$5007 line lying in the $8150$ \AA\ and
            $9140$ \AA\ filters respectively.
  \label{redshift_hist}}
  \end{centering}
  \end{figure}

The narrow-band emission-line selection produces a mixture of objects
corresponding to H$\alpha$, \oiii$\lambda$5007, and \oii$\lambda$3727
and, at the faintest magnitudes ($>24$), high redshift ($z>5$)
Ly$\alpha$ emitters.  There are almost no high redshift Lyman alpha
emitters at magnitudes brighter than 24.  The distribution of the
redshifts for objects other than the LAEs is shown in
Figure~\ref{redshift_hist}. The largest fraction of objects corresponds
to cases where the \oiii$\lambda$5007 line lies in the narrow band
filters. For the present work only the objects selected in \oiii\ or
H$\alpha$ are of interest, since we cannot determine the metallicities
or the Balmer line strengths  of the \oii\ selected samples without
futher near infrared spectroscopy.  Our final sample of galaxies
therefore consists of 214 galaxies with redshifts between zero and one
of which 189 are chosen with \oiii\ and 25 with H$\alpha$.  125 of these
have narrow band magnitudes brighter than AB=24.

  \begin{figure*}
  \begin{centering}
  \vspace*{-1.8in}
    \includegraphics[viewport=20 10 486 636,width=2.9in,angle=0,scale=1.0]{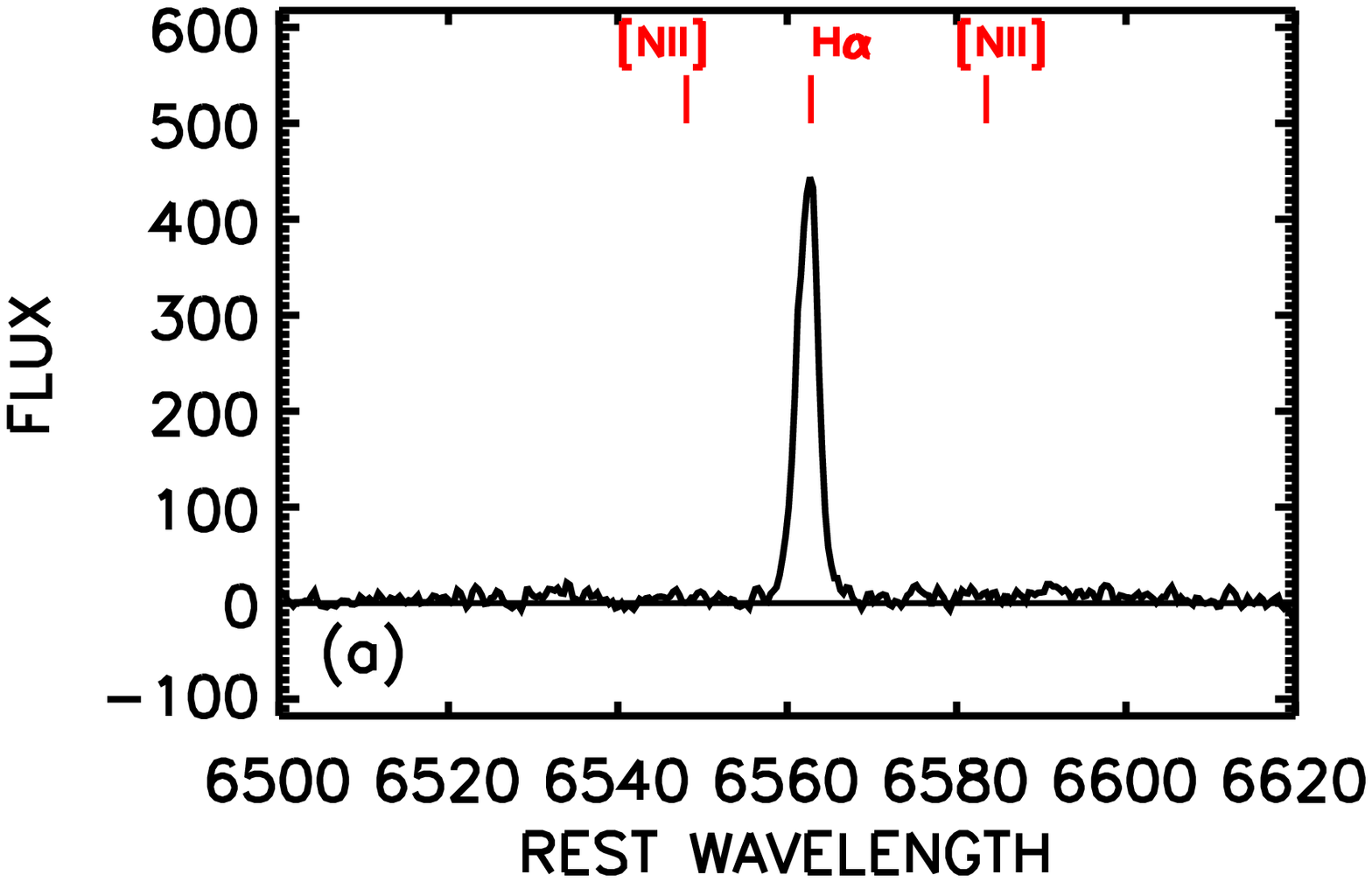}
    \includegraphics[viewport=20 10 486 636,width=2.9in,angle=0,scale=1.0]{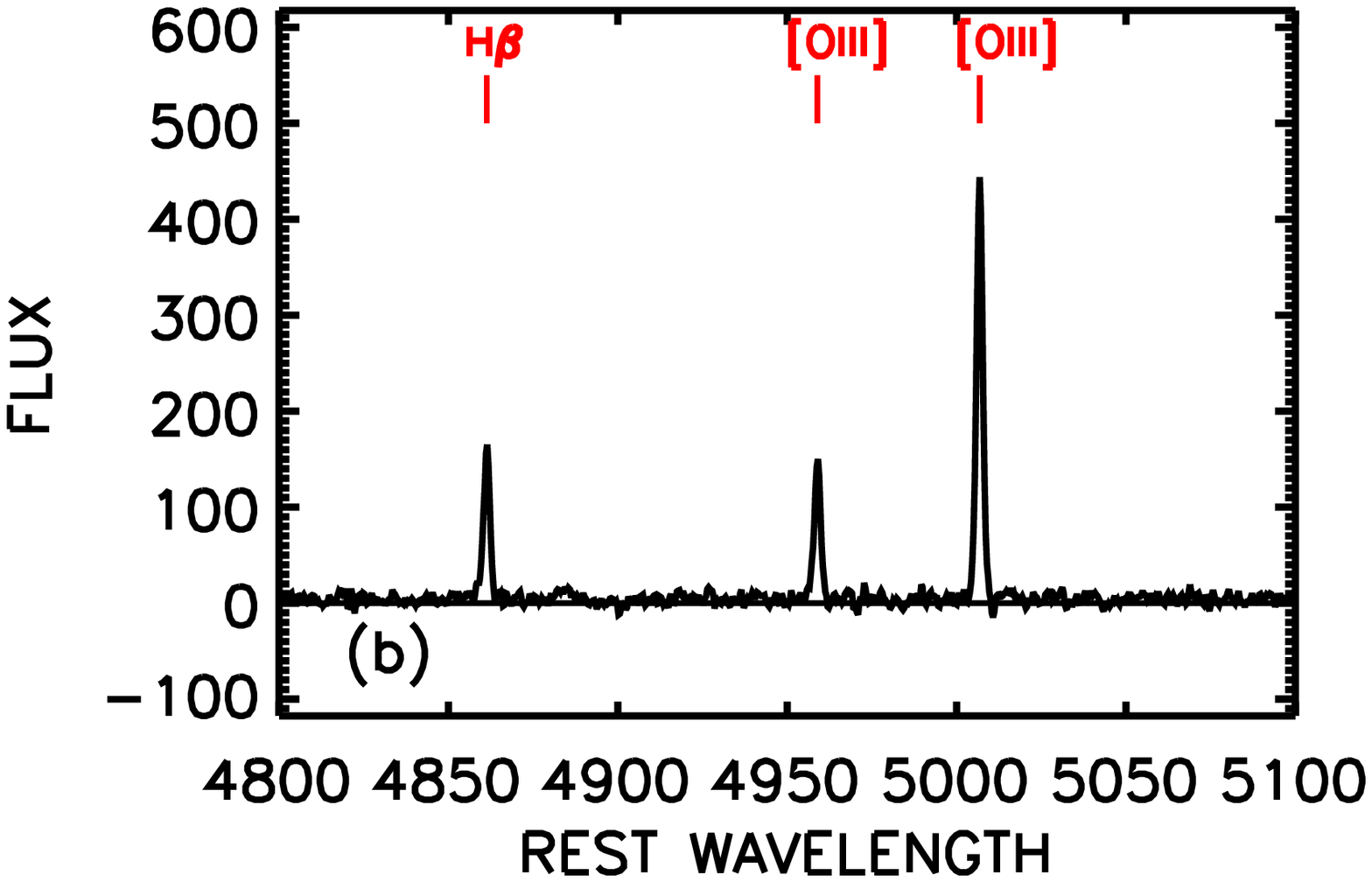}\vspace*{-1.8in}
    \includegraphics[viewport=20 10 486 636,width=2.9in,angle=0,scale=1.0]{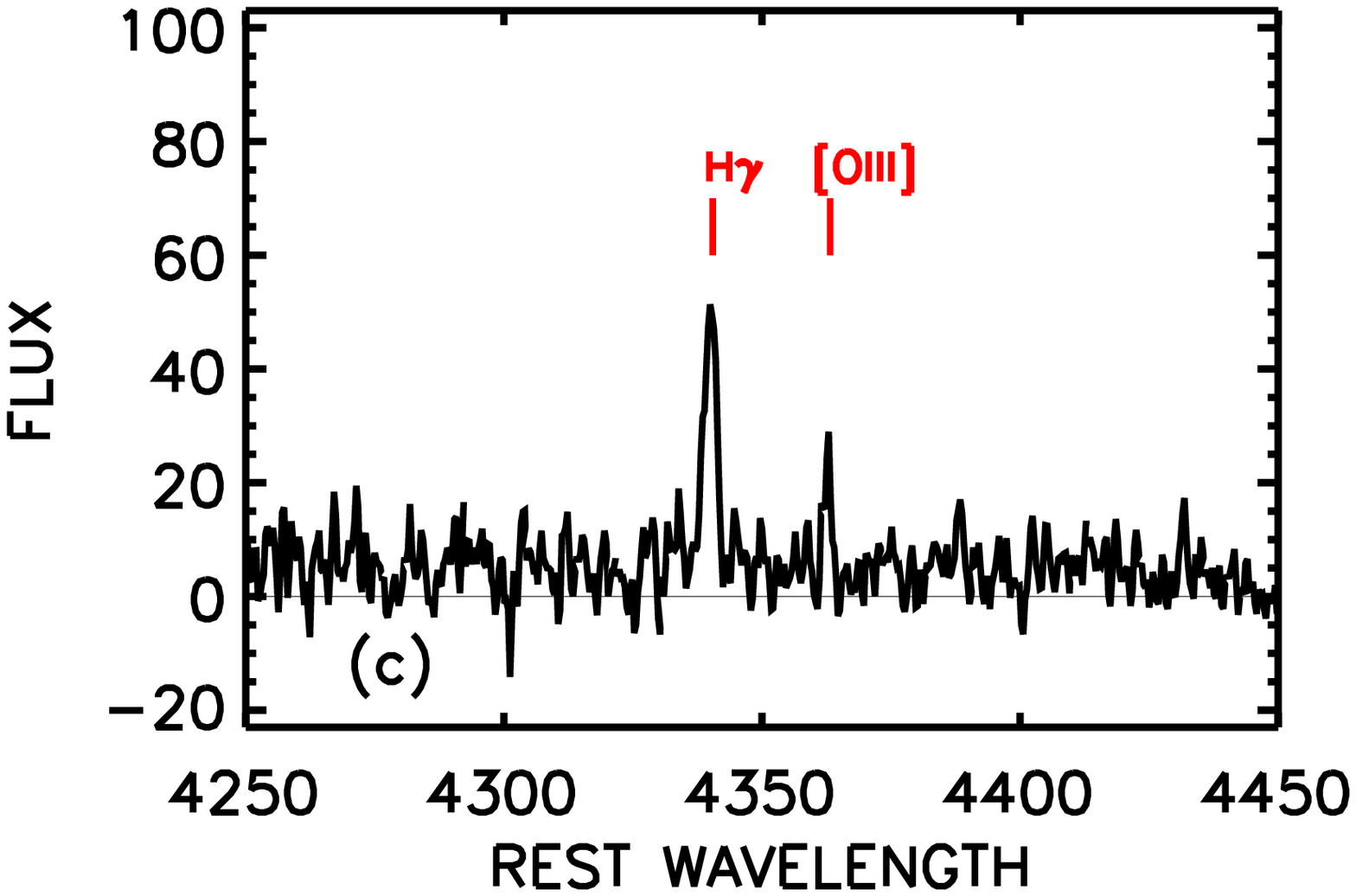}
    \includegraphics[viewport=20 10 486 636,width=2.9in,angle=0,scale=1.0]{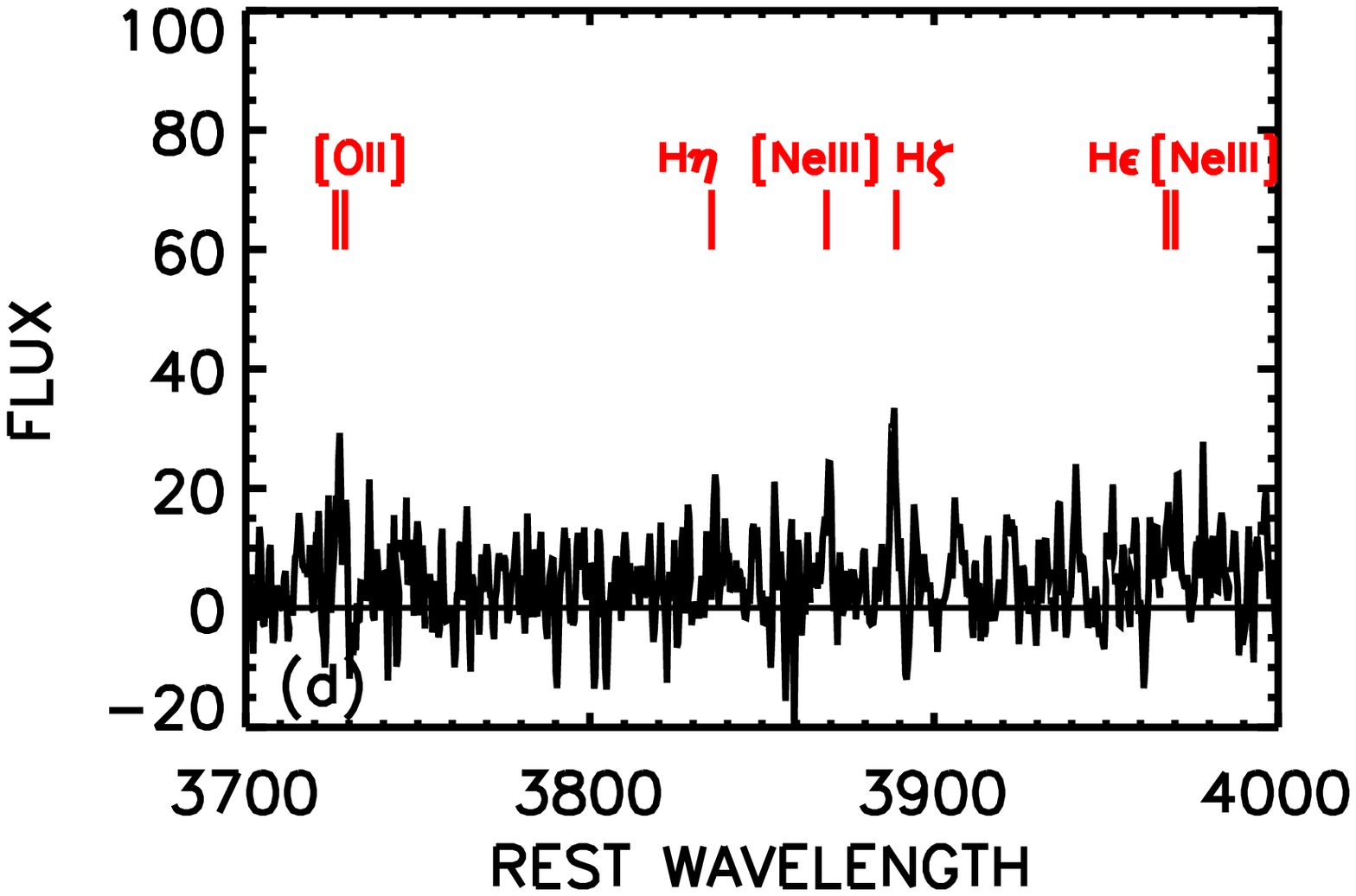}
    \caption{Portions of the spectrum of the the lowest metallicity 
             galaxy in the sample showing detected emission lines.  This 
             is object 29 in the \citet{kakazu} catalog of objects 
             selected in the NB912 filter.  It lies at a redshift of 
             0.3931.  The emission line features are labeled and marked 
             with the solid lines.  In panels (a) and (b) we have 
             increased the scale of the vertical axis to show the extremely
             strong H$\alpha$, \oiii, and H$\beta$ lines.  The \oii\ line 
             is only marginally detected in this spectrum and the two \nii\ 
             lines in panel (a) are not seen.
  \label{sample_spectrum_one}}
  \end{centering}
  \end{figure*}

\section{Flux Calibrations}
\label{secflux}

Generally our spectra were not obtained at the parallactic angle since
this is determined by the DEIMOS mask orientation required to maximize
object placement in slits over the entire mask field.  Therefore, flux
calibration using standard stars is problematic because of atmospheric
refraction effects, and special care must be taken for the flux
calibration. A much more extensive discussion can be found in
\citet{kakazu} but here we focus only on measurements of the relative
line fluxes from the spectra which we require for the metallicity
measurements.

  \begin{figure}
  \vspace*{-0.15in}
  \begin{centering}
    \includegraphics[width=3.4in,angle=0,scale=0.95]{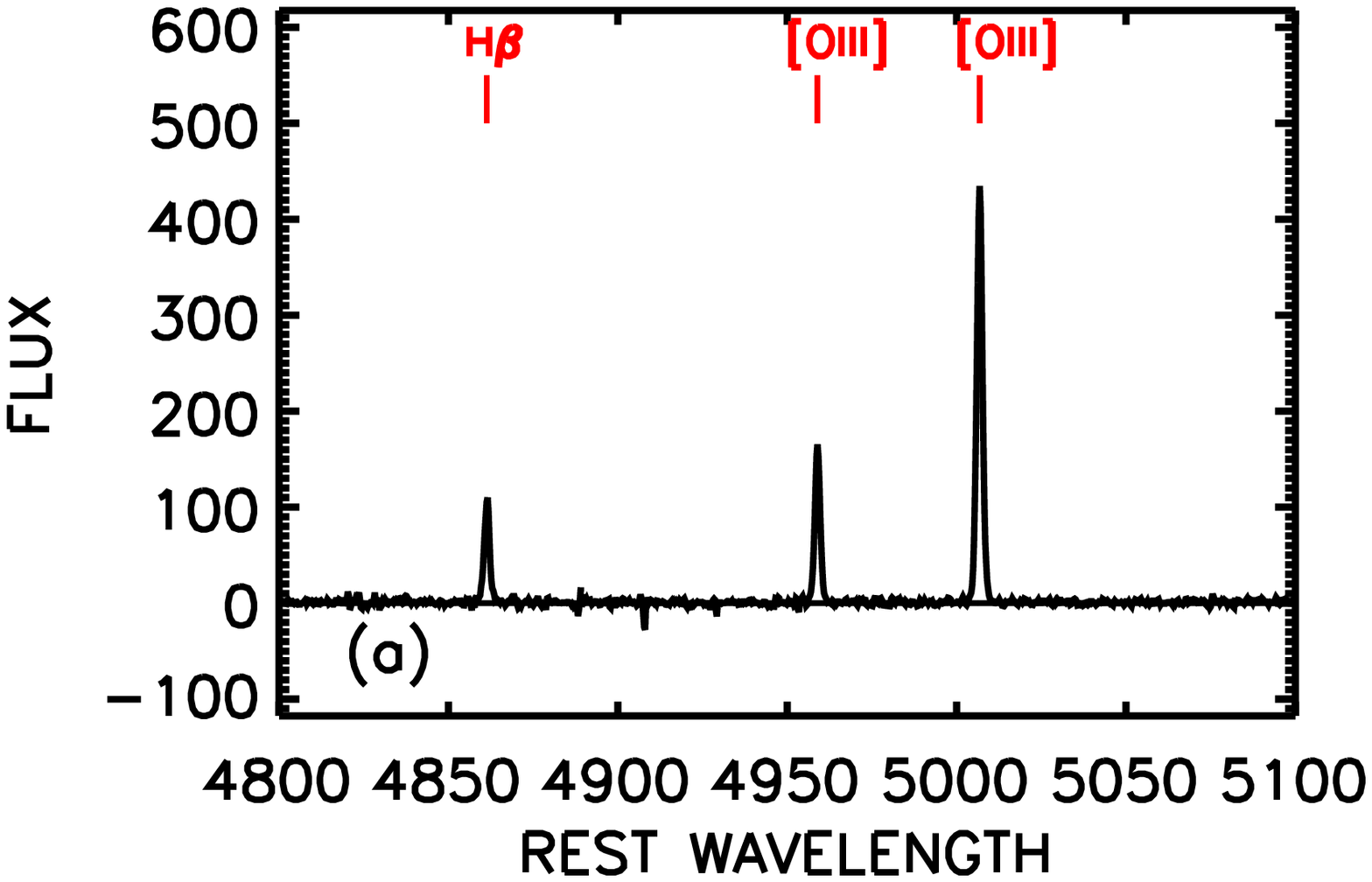}
    \includegraphics[width=3.4in,angle=0,scale=0.95]{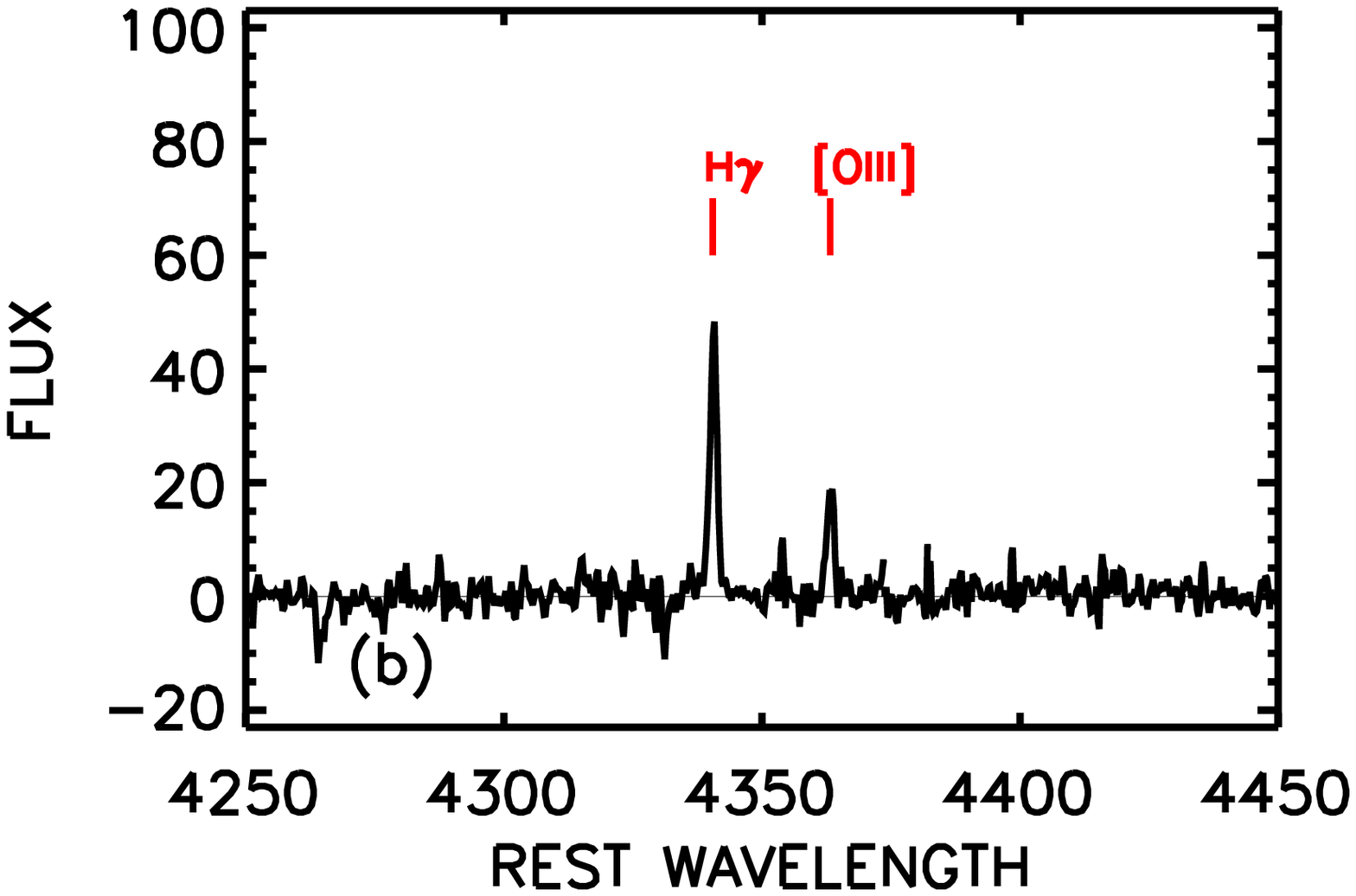}
    \includegraphics[width=3.4in,angle=0,scale=0.95]{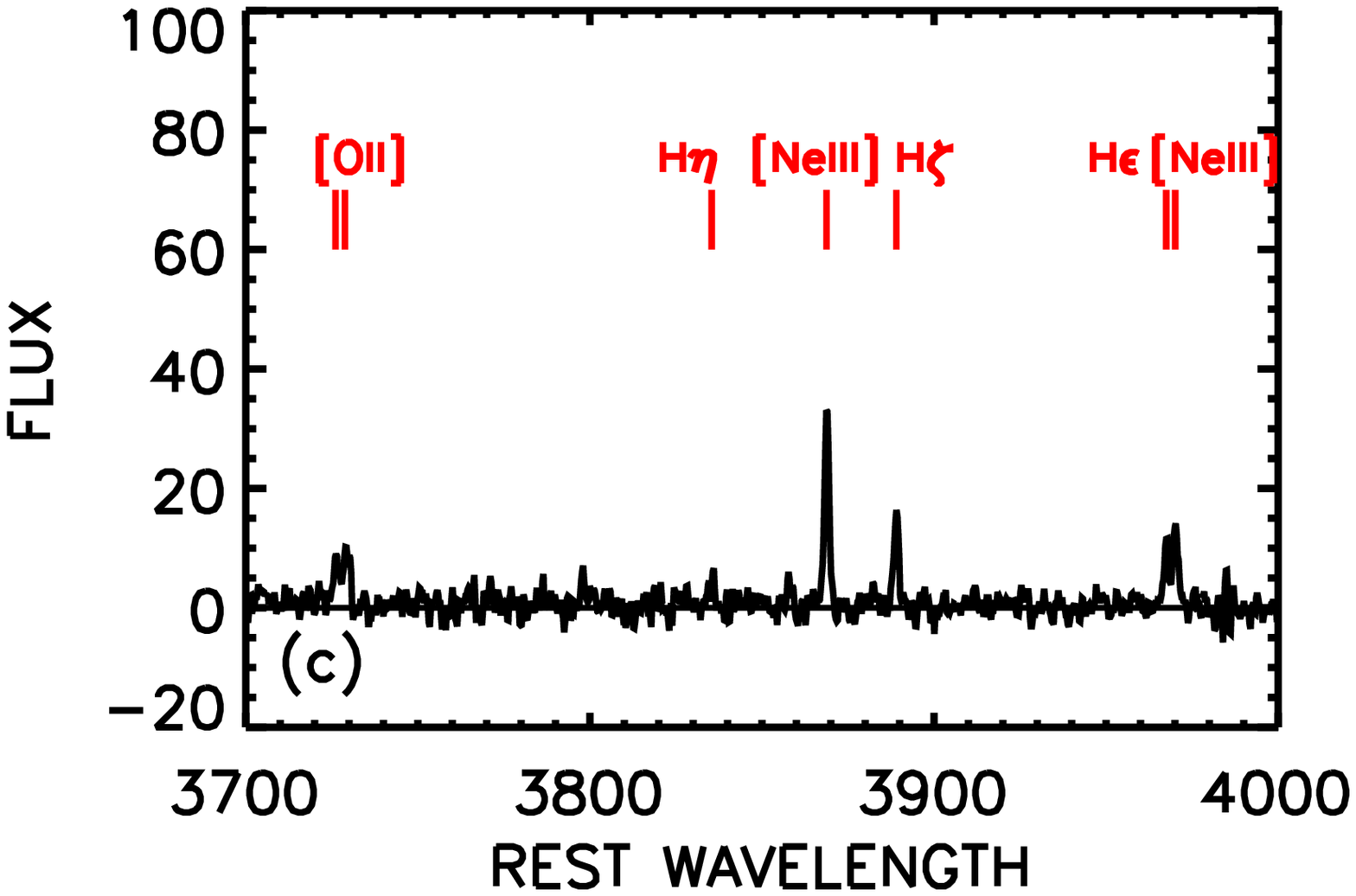}
    \caption{Portions of the spectrum of the second lowest metallicity
             galaxy in the sample showing detected emission lines. 
             This is object 269 in the \citet{kakazu} catalog of 
             objects selected in the NB912 filter.  It lies at a 
             redshift of 0.8175.  In (a) we have increased the scale 
             of the vertical axis to show the extremely strong \oiii\ 
             and H$\beta$ lines.  The emission line features are 
             labeled and marked with the solid lines.
  \label{sample_spectrum}}
  \end{centering}
  \end{figure}

The spectra were initially relatively calibrated using the measured
instrument response.  Portions of the spectra of the two lowest
metallicity galaxies in the sample are shown in
Figures~\ref{sample_spectrum_one} and \ref{sample_spectrum}.  Relative
line fluxes can be robustly measured from the spectra without flux
calibration as long as we restrict the line measurements to short
wavelength ranges where the DEIMOS response is essentially constant.  For
example, one can assume that the responses of neighboring lines (e.g.
\oiii$\lambda$4959 and \oiii$\lambda$5007) are the same and therefore one
can measure the flux ratio without calibration. The present problem is
considerably simplified by the presence of the H$\beta$ line near
\oiii$\lambda$4959 and \oiii$\lambda$5007
(Figure~\ref{sample_spectrum_one}(b) and Figure~\ref{sample_spectrum}(a))
and the H$\gamma$ line near \oiii$\lambda$4363
(Figure~\ref{sample_spectrum_one}(c) and Figure~\ref{sample_spectrum}(b)).
For these lines, and also for the \nii$\lambda$6584 line
(Figure~\ref{sample_spectrum_one}(a)) we can therefore use the Balmer line
ratios to provide extinction corrected flux ratios for the metal lines. To
calibrate these lines, we used neighboring Balmer lines with the
assumption of Case B recombination conditions. We cannot so easily do this
near \oii$\lambda$3727 where the Balmer lines are weak and in some cases
contaminated (Figure~\ref{sample_spectrum_one}(d) and
Figure~\ref{sample_spectrum}(c)).  Fortunately the \oii$\lambda$3727 is
generally extremely weak in the spectra and the uncertainty in the
calibration has little effect on the metal determination.

For each spectrum we fitted a standard set of lines. For the stronger
lines we used a full Gaussian fit together with a linear fit to the
continuum baseline.  For weaker lines we held the full width constant
using the value measured in the stronger lines and set the central
wavelength to the nominal redshifted value. The \oii$\lambda$3727 line
was fitted with two Gaussians with the appropriate wavelength
separation.  We also measured the noise as a function of wavelength by
fitting to random positions in the spectrum and computing the dispersion
in the results.

\begin{figure}
  \vspace*{-0.2in}
  \begin{centering}
    \includegraphics[width=3.5in,angle=0,scale=0.95]{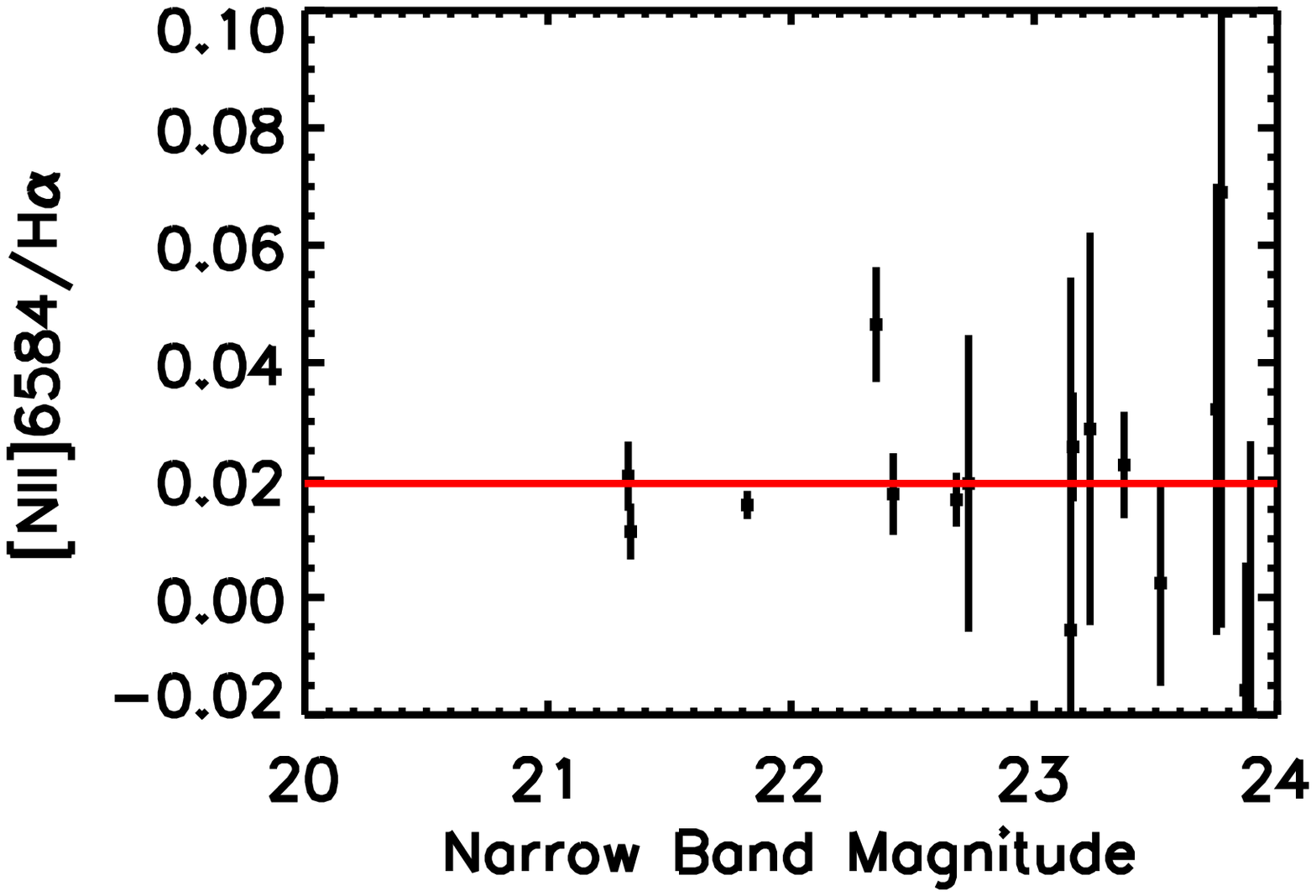}
    \includegraphics[width=3.5in,angle=0,scale=0.95]{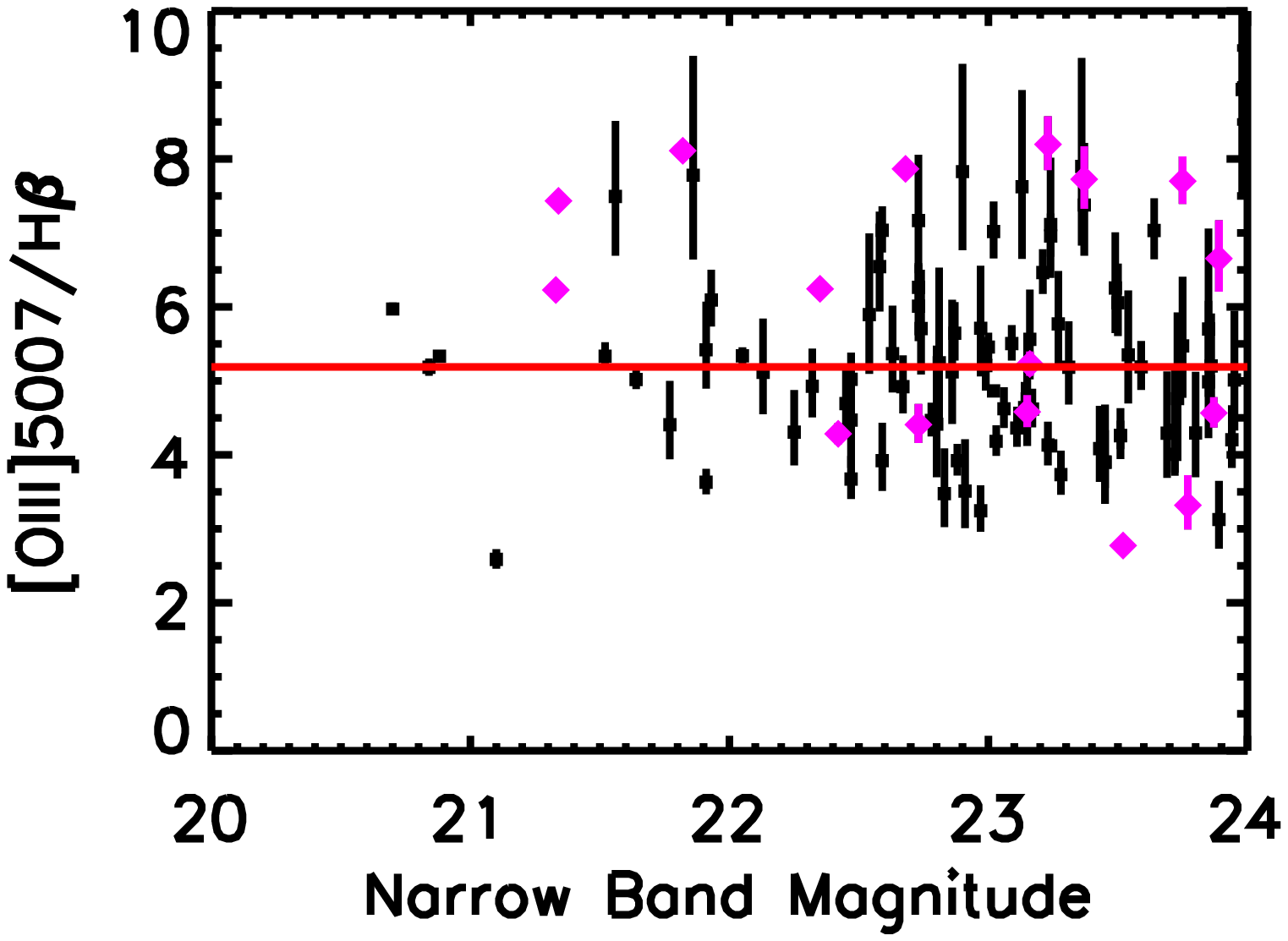}
    \includegraphics[width=3.5in,angle=0,scale=0.95]{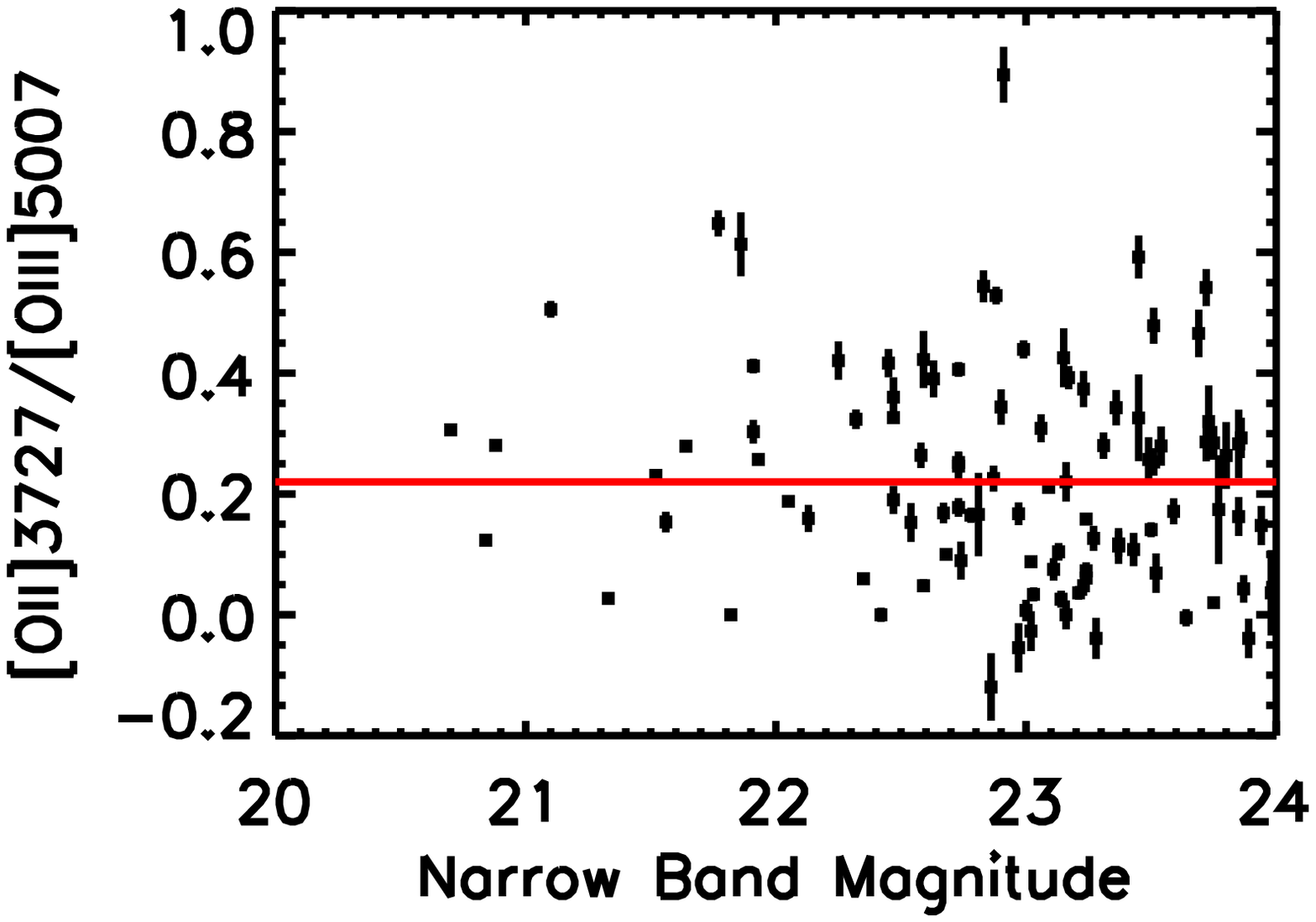}
    \caption{Flux ratios of \nii$\lambda$6584/H$\alpha$ (panel a), 
             \oiii$\lambda$5007/H$\beta$ (b), and 
             \oii$\lambda$3727/\oiii$\lambda$5007 (c) computed from the 
             spectra. The line ratios are only plotted for spectra
             where the signal to noise of the H$\beta$ line is greater 
             than five.  In a small number of cases the line fluxes of 
             the weak \nii$\lambda$6584 and \oii$\lambda$3727 lines in 
             the spectra scatter to negative values. The data are plotted 
             against the narrow band magnitude and the error bars are one 
             sigma.  \protect{\ha} selected spectra are shown with purple 
             diamonds in panel (b).  The red solid curves show the median 
             values in the sample.
  \label{flux-ratios}}
  \end{centering}
  \end{figure}

The \nii$\lambda$6584/H$\alpha$, \oiii$\lambda$5007/H$\beta$, and
\oii$\lambda$3727/\oiii$\lambda$5007 ratios are shown as a function of the
narrow band magnitude in the selection filter in Figure~\ref{flux-ratios}.
The population is relatively uniform in its line properties. Nearly all of
the galaxies have weak \nii/H$\alpha$, (a median ratio of 0.02), weak
\oii$\lambda$3727 relative to \oiii$\lambda$5007, (a median ratio of 0.22),
and strong \oiii$\lambda$5007/H$\beta$, (a median ratio of 5.2). There
seems to be little difference between the H$\alpha$ selected population
(shown with purple diamonds in Figure~\ref{flux-ratios}b) and the \oiii\
selected population (shown with black squares). The very weak \nii\ lines,
in combination with the very strong \oiii, show that these galaxies are not
excited by active galactic nuclei, and that the galaxies have very high
ionization parameters and low metallicity, as we shall quantify in the next
section.

\section{Galaxy Metallicities}
\label{secmet}

The spectra are of variable quality, reflecting the range of magnitudes
and exposure times, and, in order to measure the metallicities, we need
very high signal to noise observations.  It is also important that the
Balmer lines are well detected since our flux calibrations rely on  the
neighboring Balmer lines.  We therefore restrict ourselves to emitters
whose H$\gamma$ line fluxes are detected above the ten sigma level.
Among the 25 H$\alpha$  selected emitters in our total spectroscopic
sample there are 8 sources with spectra of sufficient quality for the
metallicity analysis, and among the 189 sources selected using
\oiii$\lambda$5007 there are 23 such sources, giving a total sample of
31 sources for the metallicity analysis.

\begin{figure}
  \begin{centering}
    \includegraphics[width=3.4in,angle=0,scale=1.]{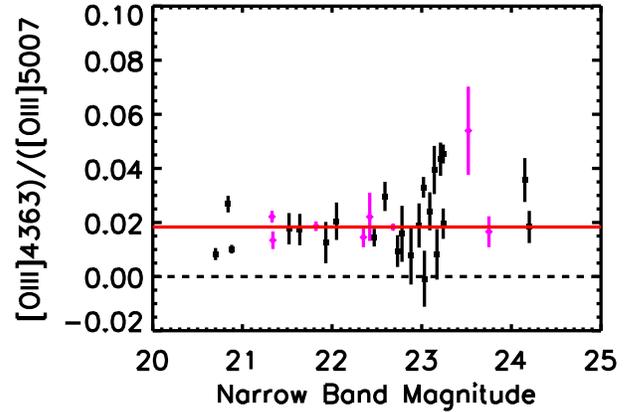}
    \caption{Ratio of the \oiii$\lambda$4363/\oiii$\lambda$5007 
             fluxes.  The data are plotted against the narrow band
             magnitude and the error bars are one sigma. Only spectra
             where the signal to noise of the H$\gamma$ line is above 
             ten are included. The values for the H$\alpha$ selected 
             spectra are shown with purple diamonds and for the 
             \oiii$\lambda$5007 selected spectra with black squares.
             The red solid line show the median value in the combined 
             sample of 31 galaxies. The black solid line shows the zero 
             ratio. Only three of the sources do not have 
             \oiii$\lambda$4363 detected above the one sigma level.
  \label{o3t_plot}}
  \end{centering}
  \end{figure}

As we illustrate in Figure~\ref{o3t_plot}, nearly all of these sources
are detected in the \oiii$\lambda$4363 auroral line.  Of the 31 sources,
23 are detected above the 3 sigma level and 9 above the 5 sigma level.
The median value of the ratio of the \oiii$\lambda$4363 auroral line to
the \oiii$\lambda$5007 line is 0.018 for the 31 objects and there is no
indication that the values seen in spectra selected with the H$\alpha$
line, which are shown by the purple diamonds in Figure~\ref{o3t_plot},
are any different from those selected with \oiii$\lambda$5007 which are
shown by the black squares.

The presence of \oiii$\lambda$4363, immediately suggests that these are
metal-deficient systems but, more importantly, it allows us to determine
the electron temperature from the ratio of the \oiii$\lambda$4363 line
to \oiii$\lambda$$\lambda$5007,4959.  This procedure is often referred
to as the `direct' method or $T_e$ method
\citep[e.g.,][]{seaton,pagel92,pilyugin,izo06c}.  To derive $T_e$\oiii\
and the oxygen abundances, we used the \citet{izo06c} formulae, which
were developed with the latest atomic data and photoionization models.
Using the \citet{pagel92} calibrations with the $T_e$\oii$-$$T_e$\oiii\
relations derived by \citet{garnett92}, gives consistent abundances
within 0.1 dex.  The \sii$\lambda\lambda$6717, 6731 lines that are
usually used for the determination of the electron number density, are
beyond the Keck/DEIMOS wavelength coverage for our \oiii\ emitters.
Therefore we assumed n$_e$ = 100 cm$^{-3}$.  However the choice of
electron density has little effect as electron temperature is
insensitive to the electron density; indeed we get the same results even
when we use n$_e$ = 1,000 cm$^{-3}$.

The spectra of the three lowest redshift H$\alpha$ emission-line
selected galaxies do not cover the \oii$\lambda$3727 line, leaving us
with a final sample of 28 galaxies for our analysis. Seven of these
objects satisfy the definition of XMPGs [12 + log(O/H) $<$ 7.65;
\citealt{kunth}].  The lowest metallicity galaxes in the sample are
KHC912-29 with 12 + log(O/H) $= 6.97\pm0.17$, and KHC912-269 which has
12 + log(O/H) $= 7.25\pm0.03$. The spectra of these two galaxies are
shown in Figures~\ref{sample_spectrum_one} and \ref{sample_spectrum}.
Their metallicities are comparable to the currently known most
metal-poor galaxies [I Zw 18 and SBS0335$-$052W; 12 + log(O/H) $\sim$
7.1 $-$ 7.2].

\section{Discussion}
\label{secdisc}

\subsection{Metallicity and the ionization parameter}
\label{secmet_ion}

\begin{figure}
  \begin{centering}
    \includegraphics[width=3.4in,angle=0,scale=1]{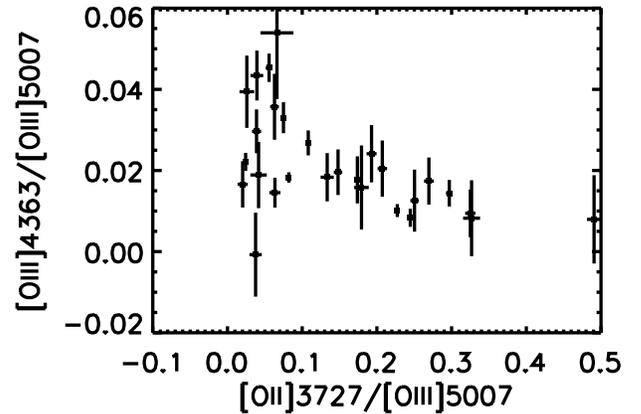}
    \caption{\protect{\oiii}$\lambda$4959$+$$\lambda$5007/\protect{\oiii}$\lambda$4363
             versus \protect{\oii}$\lambda$3727/\protect{\oiii}$\lambda$5007
	     for the 28 galaxies in the metallicity sample. One sigma 
             error bars are shown for both ratios.
  \label{o_ratios}}
  \end{centering}
  \end{figure}

Figure~\ref{o_ratios} shows the electron temperature sensitive line
ratio, \oiii($\lambda$4363)/\oiii$\lambda$5007 versus
\oii$\lambda$3727/\oiii$\lambda$5007.  If we have an estimate of the
metallicity, as in the present case, we can use the
\oii$\lambda$3727/\oiii$\lambda$5007 ratio to estimate the ionization
parameter $q$, defined as the number of hydrogen ionizing photons
passing through a unit area per second per unit hydrogen number density.
We can see from Figure~\ref{o_ratios} that there is a strong inverse
correlation of \oiii$\lambda$4363 and \oii$\lambda$3727. Systems with
weak \oii$\lambda$3727 generally have strong \oiii$\lambda$4363.

\begin{figure}
  \vspace*{-0.2in}
  \begin{centering}
    \includegraphics[width=3.4in,angle=0,scale=1]{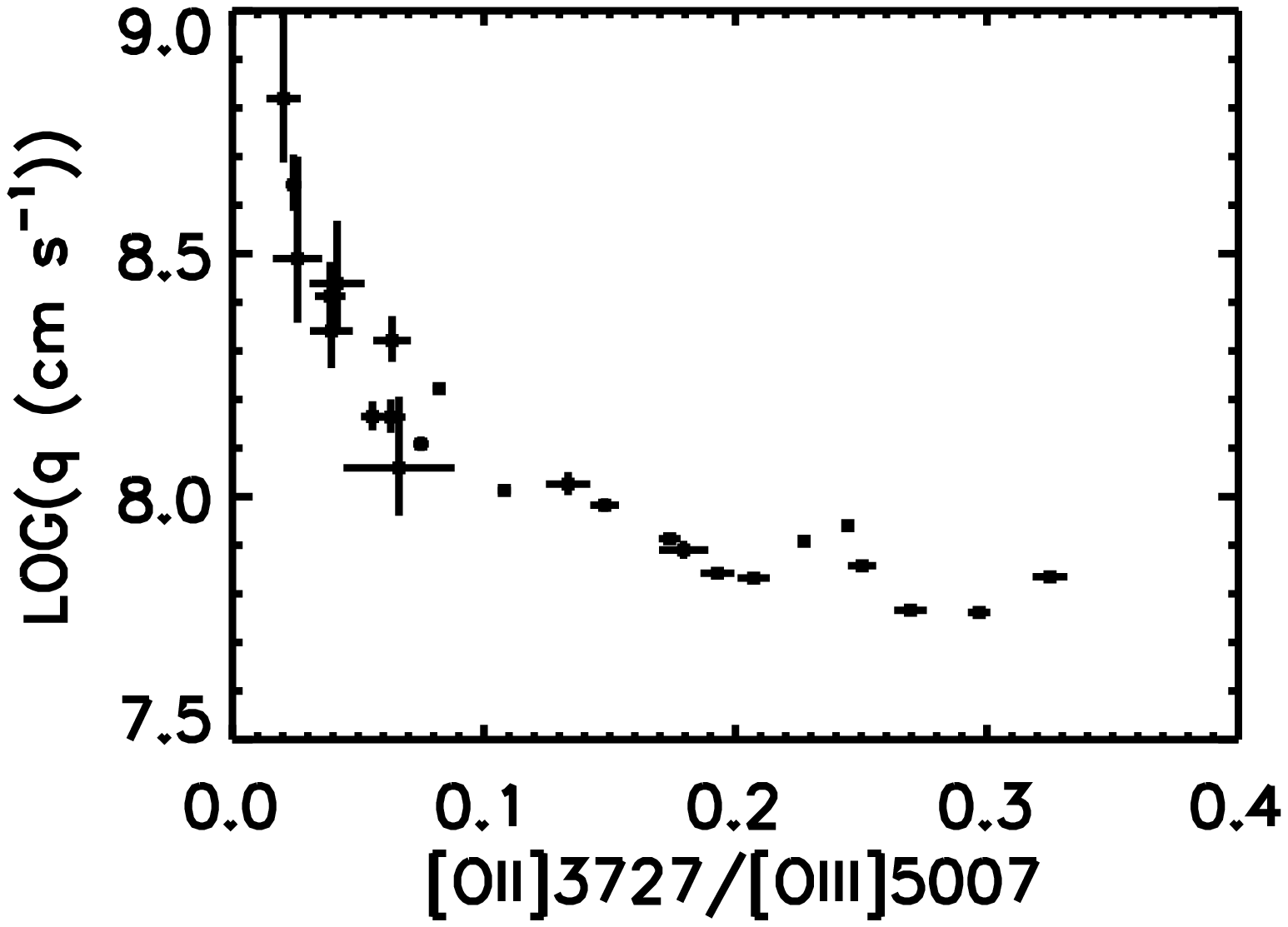}
    \includegraphics[width=3.4in,angle=0,scale=1]{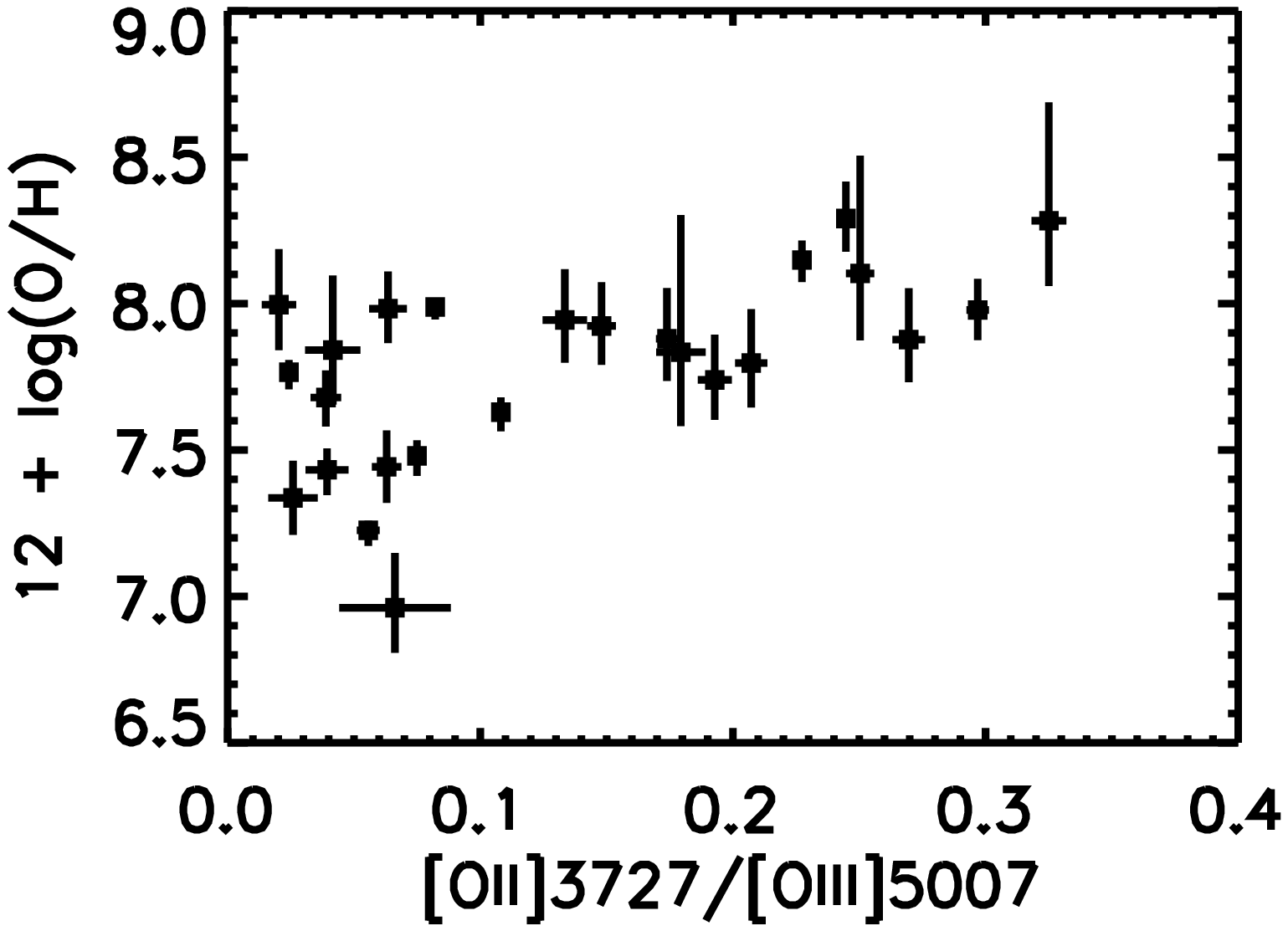}
    \caption{(a) The ionization parameter $q$ versus  
             \protect{\oii}$\lambda$3727/\protect{\oiii}$\lambda$5007 
             for the metallicity sample. (b) 12 +log(O/H) versus             
             \protect{\oii}$\lambda$3727/\protect{\oiii}$\lambda$5007 
             for the metallicity sample.  Only the 25 objects with 
             \oiii$\lambda$4363 detected above the one sigma level 
             are shown and one sigma error bars are given in both cases.
  \label{o32_plots}}
  \end{centering}
  \end{figure}

We have computed the ionization parameters ($q$) for the sample using
the parameterized forms of the dependence of
\oii$\lambda$3727/\oiii$\lambda$5007 on $q$ and 12+log(O/H) from
\citet{KK04}. In Figure~\ref{o32_plots}(a) we show the dependence of the
$q$ parameter on \oii$\lambda$3727/\oiii$\lambda$5007 and in
Figure~\ref{o32_plots}b we show the dependence of 12+log(O/H) on
\oii$\lambda$3727/\oiii$\lambda$4363.

Their is a clear dependence between the metallicity and
\oii$\lambda$3727/\oiii$\lambda$5007. Objects with low metallicity have
low \oii$\lambda$3727/\oiii$\lambda$5007 and all of the XMPGs have
values of this ratio below 0.12.  Conversely, seven of the thirteen
galaxies satisfying this condition are XMPGs.  This is potentially a
very valuable tool for optimizing the search for the lowest metallicity
galaxies at these redshifts since we can focus our spectrosopic followup
on galaxies with weak \oii$\lambda$3727 lines.

The ionization parameter also increases with decreasing
\oii$\lambda$3727/\oiii$\lambda$5007 (Figure~\ref{o32_plots}(a)) though
the ionization parameters lie in a relatively narrow range for
log(q)$\sim$7.8-8.5.  However, there is a considerable variation in the
theoretical models \citep[e.g.,][]{mcgaugh91,pilyugin00} and the exact
shape and normalization are somewhat uncertain. The range of ionization
parameters is similar to the values found in magnitude limited samples
of galaxies at the same redshift \citep{cowie08}.

\begin{figure} [t]
  \begin{centering}
    \includegraphics[width=3.4in,angle=0,scale=1]{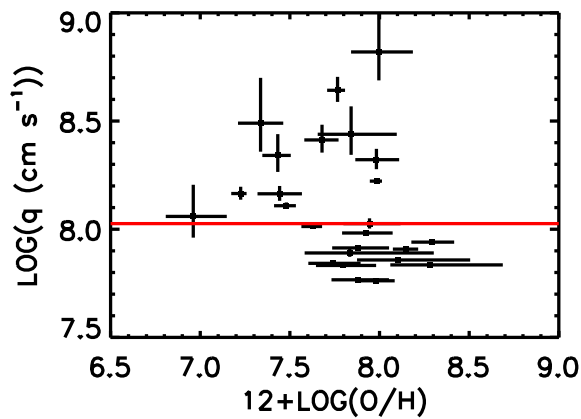}
    \caption{The $q$ parameter versus oxygen abundance
             for the metallicity sample.  
             One sigma errors are shown for the oxygen abundances and
             the $q$ parameters. Only the 25 objects with greater than
             one sigma detections of the \oiii$\lambda$4363 line are shown.
             The red line shows the median value of the $q$ parameter
             in the sample.
  \label{o_q}}
  \end{centering}
  \end{figure}

In Figure~\ref{o_q} we show the ionization parameter q as a function of
the metallicity. It is clear that the lower metallicity galaxies have
higher ionization parameters. However, we again note the uncertainties
in the theoretical modelling.

\subsection{Metallicity and the R23 parameter}
\label{secmet_r23}

The R23 ratio [$f($\oiii$~\lambda 4959)+f($\oiii$~\lambda
5007)+f($\oii$~\lambda 3727)]/f($\hb) of Pagel et al.\ (1979) is one of
the most frequently used metallicity diagnostics.  As is well known it
is unfortunately multivalued with both a low metallicity and a high
metallicity branch. Moreover, while for the present galaxies we may be
reasonably secure that we are on the low metallicity branch
($12+\log({\rm O/H})\lesssim 8.4$), R23 is only weakly dependent on
metallicity on this branch and has a strong ionization dependence
\citep[e.g.,][]{mcgaugh91}.

Nevertheless, recent analyses of substantial samples of local galaxies
with well determined abundances from the direct method have shown a good
empirical correlation of 12+log(O/H) with R23 in the low metallicity
range \citep{nagao06,yin07}.  \citet{izo07} favor the \citeauthor{yin07}
parameterization 12+log(O/H) $= 6.486 +1.401 \times$ log(R23) based on
comparisons with their local sample. However, in comparing these to the
present sample we must be concerned about differences in the galaxy
properties and in particular whether evolution in the  distribution of
the ionization parameter between the local and distant samples might
change the relation.

\begin{figure}
  \vspace*{-0.2in}
  \begin{centering}
    \includegraphics[width=3.4in,clip,angle=0,scale=1]{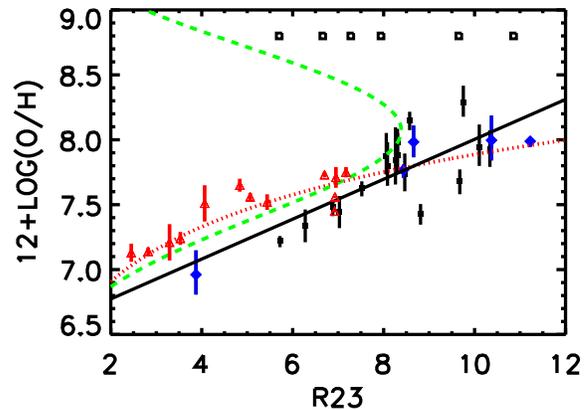}
    \caption{The oxygen  abundance versus the R23 parameter for the 
             \oiii\ selected sample (black squares) and the H$\alpha$ 
             selected objects (blue diamonds).  Galaxies with 
             \oiii$\lambda$4363 lines below the two sigma level are 
             shown with open squares at a nominal O abundance.
             One sigma errors are shown for the oxygen abundances.
             A linear fit to the combined data sets is shown by the 
             black line while fits to the local data by \citet{nagao06} 
             are shown with the green dashed line and by \citet{yin07} 
             as the red dotted line.  Local data from \citet{izo07} are 
             shown by the red triangles.
  \label{r23_o}}
  \end{centering}
  \end{figure}

In Figure~\ref{r23_o} we show the dependence of the present measurements
of 12+log(O/H) on the R23 parameter. We also show the local points from
\citet{izo07} and the empirical fits of \citet{yin07} (red dotted curve)
and \citet{nagao06}  (green dashed curve). It is clear that the present
data have lower 12+log(O/H) determinations at the same R23. This would
be expected if the ionization parameters are higher in the present
sample. A factor of five increase in the $q$ parameter could easily
produce the offsets seen.  Empirically we find a relation of the form
12+log(O/H)=6.45 + $0.15$ R23, which is shown as the black line in
Figure~\ref{r23_o}, provides a reasonable description of the data.

\subsection{Metallicity versus H$\beta$ equivalent width}
\label{secmet_ew}

As is well known the H$\beta$ equivalent width can give a rough estimate
of the age of the star formation in a galaxy.  For a Salpeter IMF and a
constant star formation rate, EW(H$\beta$) would drop smoothly to a value
of 30\AA\ at about $10^9$ yr \citep{star99} while an instantaneous
starburst would drop below this value after about $10^7$ yr. It is
therefore of considerable interest to determine the relation between
12+log(O/H) and EW(H$\beta$) in order to find if the metal build up is a
function of the age of the galaxy.

We can determine EW(H$\beta$) in two ways. We can measure it directly
from the spectra or we can determine the EW of the line falling in the
narrow band filter from the imaging observations and then use the line
ratios to determine EW(H$\beta$). However, both methods are somewhat
problematic.  Because the continuua in the spectra are extremely weak it
can be difficult to precisely measure them. The errors here are complex
and systemic efforts can be important. Thus EW(H$\beta$) can only be
confidently measured in the highest S/N spectra.  The imaging data makes
a much deeper and more accurate measurement of the continuum but
translating this to the observed frame EW of the measured line requires
us to know the filter profile extremely accurately.  This is a
particular problem for objects whose wavelengths lie on the edges of the
filter where the response changes rapidly.  We measured EW(H$\beta$)
with both methods. For the spectral determination we restricted
ourselves to spectra with S/N$>25$ in the \oiii$\lambda$5007 line. For
the images we restricted ourselves to galaxies where the transmission in
the filter was above 80\% of the peak transmission. We took the
continuum from line free broad band colors near the narrow band (the $I$
band for the NB816 filter and the $z'$ band for the NB912 filters)
assuming that the spectrum $f_\nu$ was flat. We determined the EW of the
line in the narrow band as ($10^{-0.4(N-C)}-1)/\phi$ where $N$ is the
narrow band magnitude, $C$ is the continuum magnitude and $\phi$ is the
filter response normalized such that the integral over the wavelength is
unity. Where the line in the filter was \oiii$\lambda$5007 we converted
to EW(H$\beta$) using the observed line ratios in the spectra and where
the line in the filter was H$\alpha$ we converted to EW(H$\beta$)
assuming the Case B Balmer ratios.

  \begin{figure}
  \begin{centering}
    \includegraphics[width=3.4in]{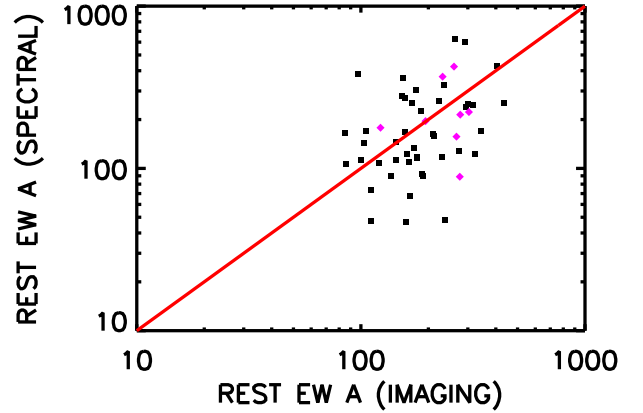}
    \caption{Comparison of rest frame equivalent widths determined from 
             the imaging data with those determined from the spectra.
             Black squares show \oiii$\lambda$5007 equivalent widths and
             purple diamonds H$\alpha$ equivalent widths. The red line 
             shows the expected relation. Only objects where the line 
             wavelength would correspond to a transmission above 80\% of 
             the peak filter response in the narrow band filter and
             where the \oiii$\lambda$5007 line is detected in the spectra 
             at a signal to noise above 25 are shown.
  \label{equivalents}}
  \end{centering}
  \end{figure}

\begin{figure}
  \vspace*{-0.2in}
  \begin{centering}
    \includegraphics[width=3.3in]{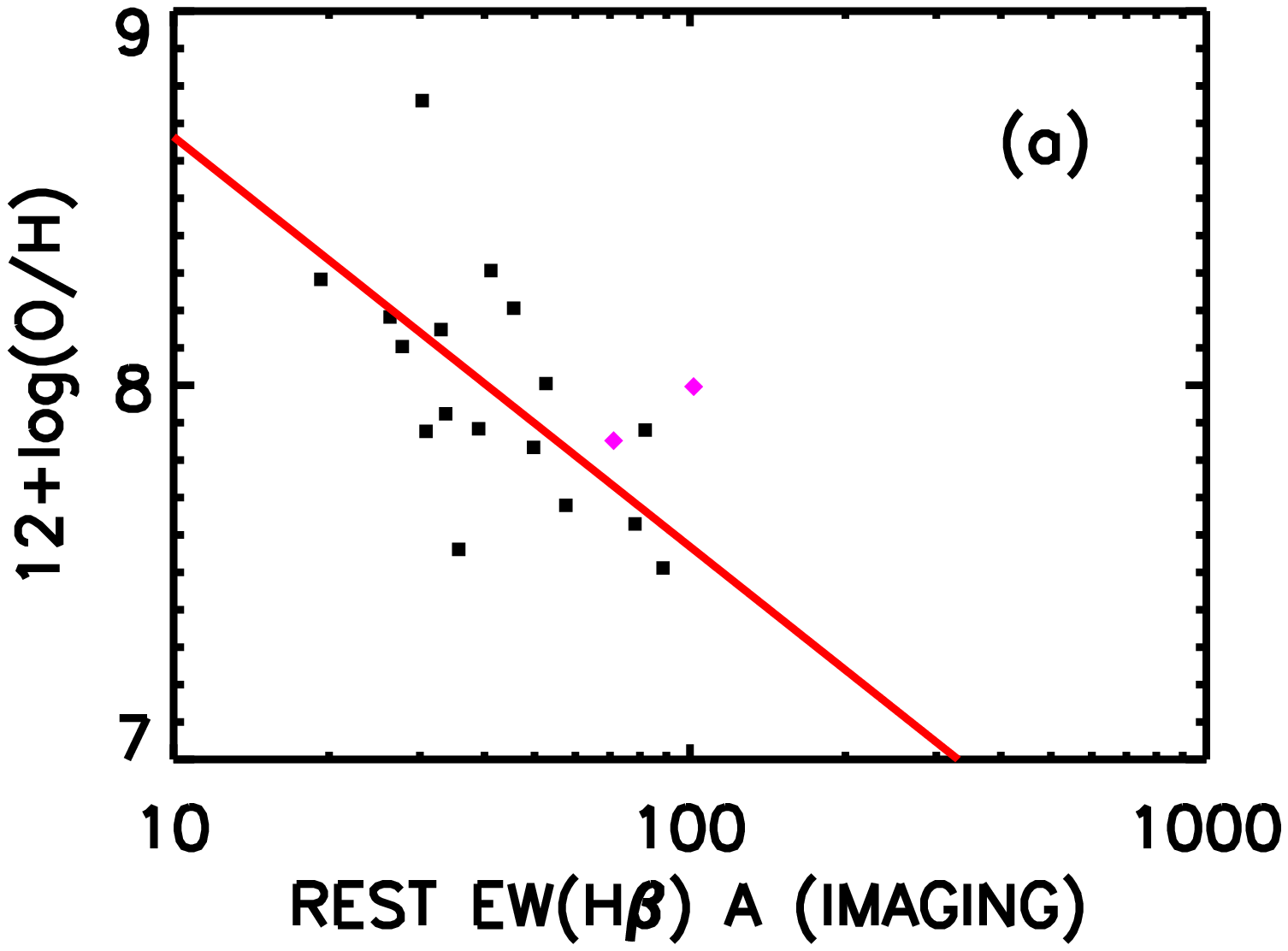}
    \includegraphics[width=3.3in]{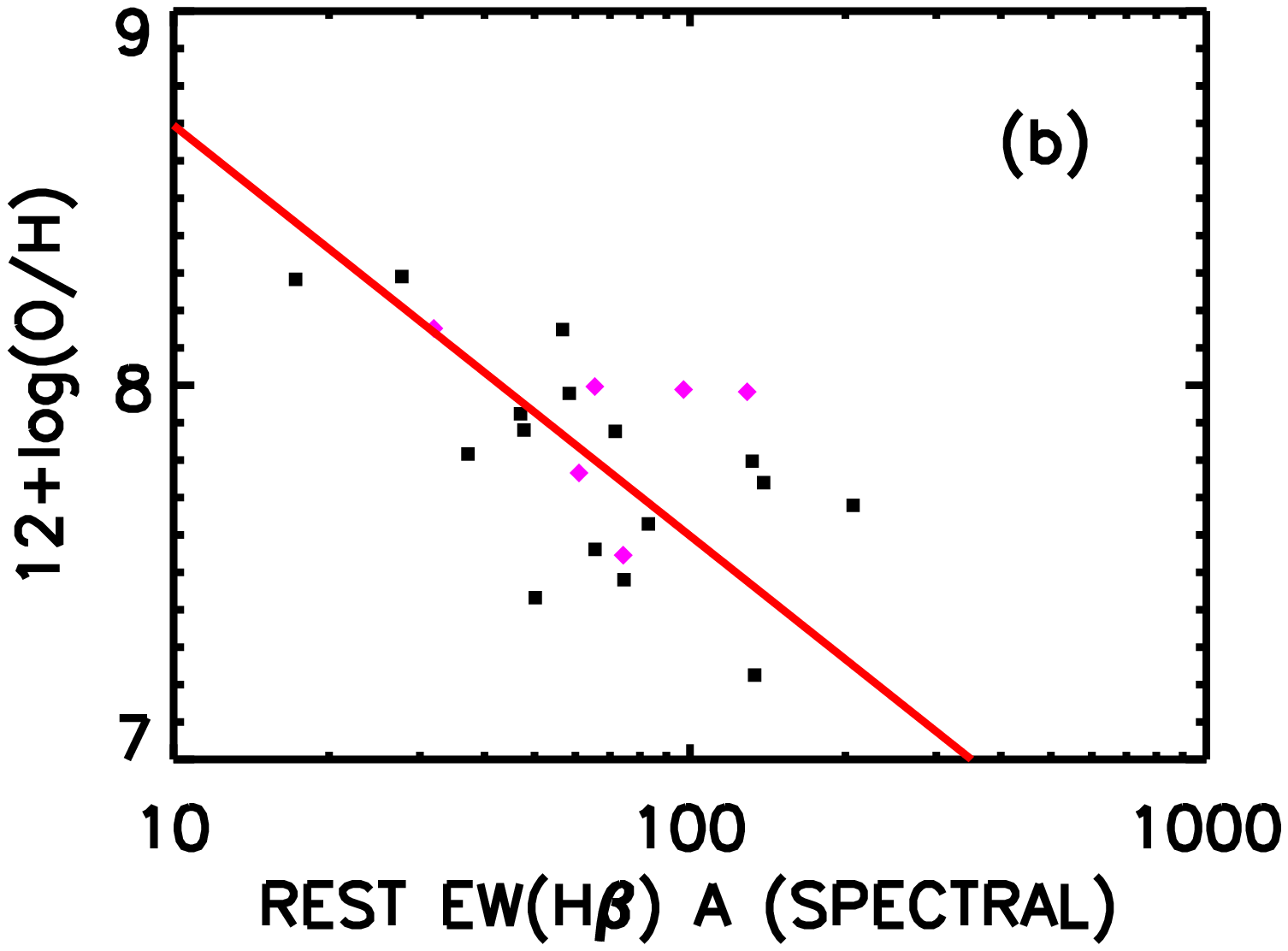}
    \caption{(a) 12+log(O/H) versus rest frame H$\beta$ equivalent 
             widths determined from the imaging data.  Black squares 
             show \oiii$\lambda$5007 selected objects and purple 
             diamonds H$\alpha$ selected objects. The red line shows 
             the best fit linear relation between 12+log(O/H) and 
             log(EW(H$\beta$)). Only objects with filter transmission 
             above 80\% of the peak and with signal to noise above 15 
             for the H$\beta$ line are shown.
             (b) Same as (a) but with equivalent widths determined
             only from the spectra. All objects with signal to noise 
             above 20 for the H$\beta$ line are shown.  The red line 
             is the fit to the imaging equivalent widths shown in (a).
  \label{o_ew}}
  \end{centering}
  \end{figure}

We compare the imaging equivalent widths with the spectroscopic
equivalent widths in Figure~\ref{equivalents}. There is broad overall
agreement but a considerable amount of scatter reflecting the
uncertainties in the two methods. However, as is shown in
Figure~\ref{o_ew}, irrespective of which method we use we see a clear
relationship  between 12+log(O/H) and EW(H$\beta$).  The best fit
relationship of 12+log(O/H) = $9.72 -1.07$ log(EW(H$\beta$) based on the
imaging equivalent widths also provides a good description for the
spectroscopically based equivalent widths.  This relation was less clear
in the equivalent widths determined from the poorer spectra used in
\citet{kakazu}. With the new data there is now clear evidence for the
build-up of metals as a function of age in the galaxies.

\subsection{Metallicity Luminosity relation}
\label{secmet_lum}

\begin{figure}
  \vspace*{-0.2in}
  \begin{centering}
    \includegraphics[width=3.4in]{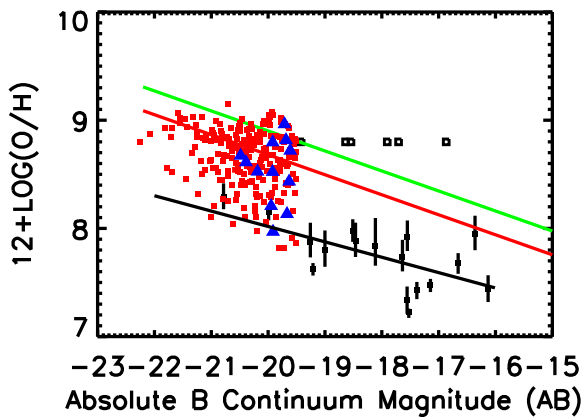}
    \caption{The oxygen  abundance versus the absolute rest frame $B$ 
             magnitude for the \oiii\ selected sample (black squares)
             The red diamonds show the oxygen abundances for a 
             magnitude limited sample in the same redshift range from
             \citet{cowie08}. The green line shows the local 
             metal-luminosity relation from \citet{trem04}, the red 
             the metallicity relation at $z=0.8$ from \citet{cowie08}, 
             and the black line the metallicity-luminosity relation of 
             the present sample.
  \label{met_lum}}
  \end{centering}
  \end{figure}

We next computed the absolute rest frame B magnitudes using magnitudes
from imaging observations in bandpasses which are clear of the emission
lines and assuming a flat $f_\nu$ spectral energy distribution to
compute the K correction.  We plot these absolute rest frame $B$
magnitudes versus the oxygen abundance derived by the direct method in
Figure~\ref{met_lum}.  The metallicity-luminosity relation can be
written as 12+log(O/H) $= 8.01 -0.15 \times (M_B(AB)-20)$.

We can compare the metallicity luminosity relation with the local
metallicity luminosity relation from \citet{trem04} (the green line in
Figure~\ref{met_lum}) and with observations of magnitude limited samples
at $z=0.6-0.9$ from \citet{cowie08} (red points) and the corresponding
metal-luminosity relation at this redshift (red line). The present
sample lies about 0.6 dex below the $z=0.6-0.9$ relation and about 0.8
dex lower than the local relation. The Cowie and Barger metallicities
are based on the upper branch of the O versus R23 relation and may miss
some low metallicity objects. We have also searched their magnitude
limited sample for strong emitters (the blue diamonds in
Figure~\ref{met_lum}) and checked these galaxies for \oiii$\lambda$4363
but none of these show the auroral line.

\section{Luminosity functions and the local Lyman alpha emitter population}
\label{lf}

\begin{figure}
  \vspace*{-0.2in}
  \begin{centering}
    \includegraphics[width=3.2in]{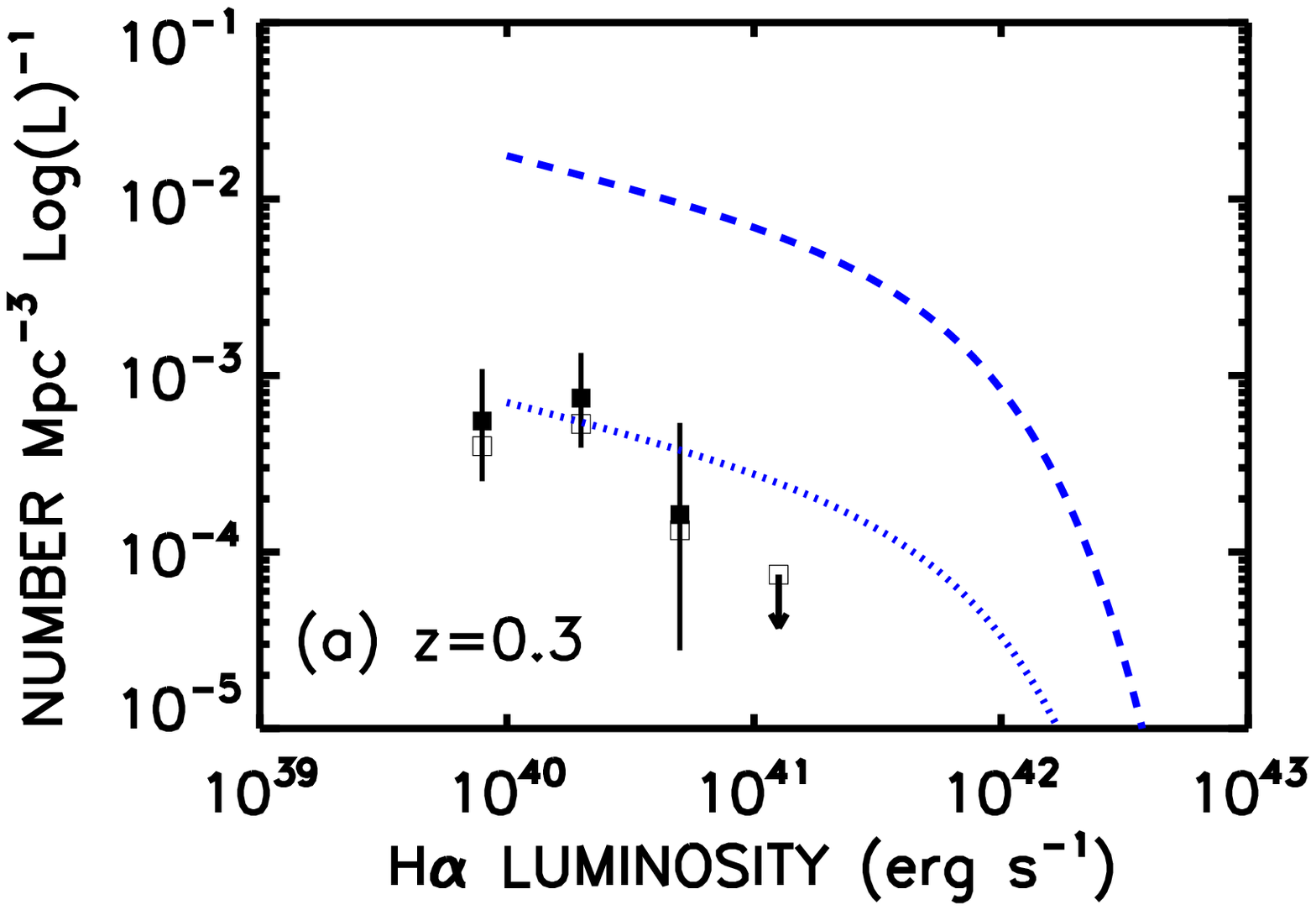}
    \includegraphics[width=3.2in]{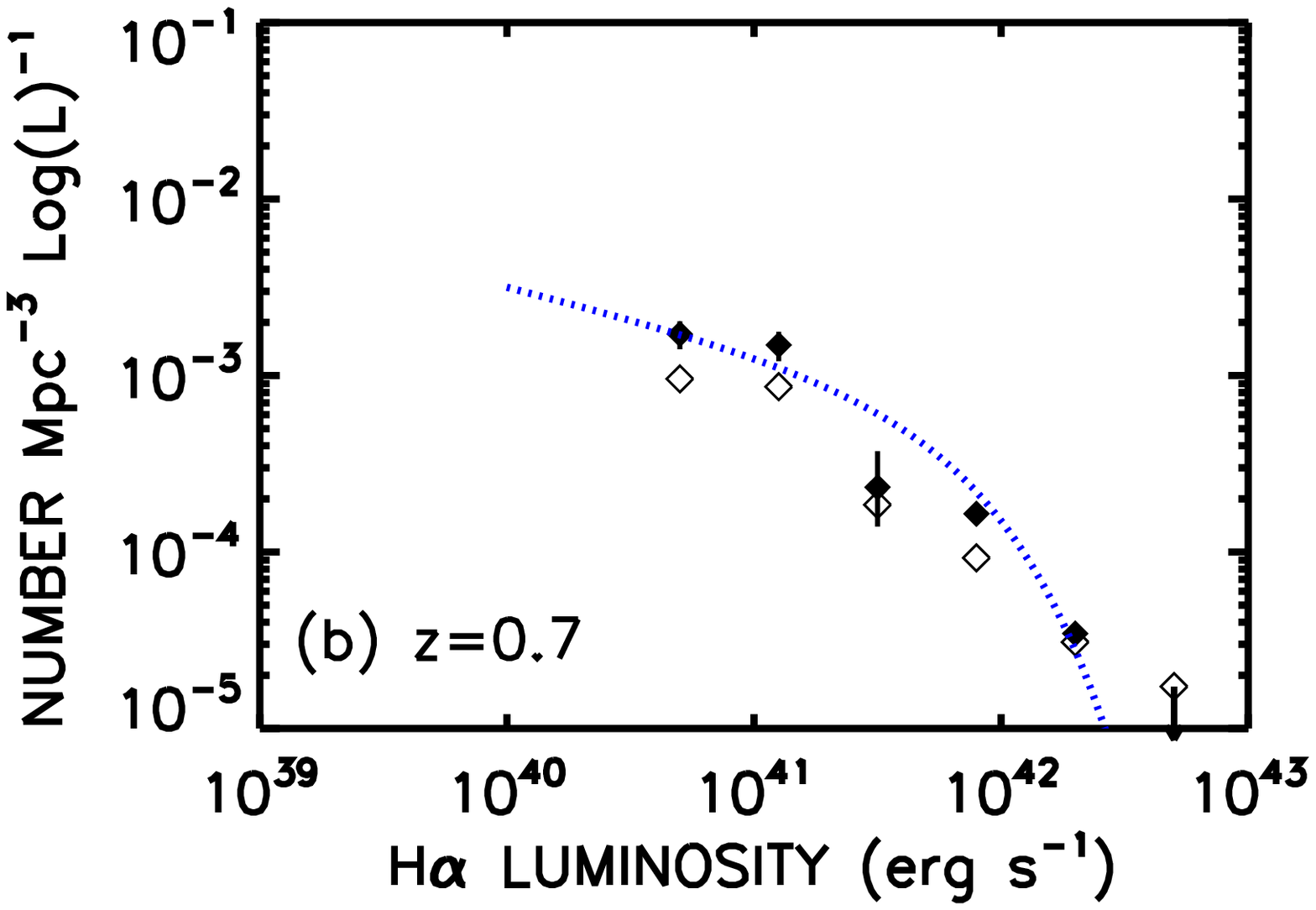}
    \caption{The luminosity function of \protect{\ha} at $z=0.3$ (top 
             panel a) and at $z=0.7$ (bottom panel b). In each case 
             the open symbols show the luminosity functions determined 
             from the spectroscopic sample alone while the solid symbols 
             show the function corrected for the incompleteness in the
             spectroscopic identification. The errors are plus and minus
             1 sigma and at the highest luminosity we show the 1 sigma
             upper limit. In panel (a) we also show the $z=0.3$ H$\alpha$ 
             luminosity function from \citet{tresse98} as the blue dashed 
             line. The dotted blue line shows the Tresse and Maddox 
             function multiplied by 0.04 which approximately matches the 
             USEL H$\alpha$ luminosity function. In panel (b) we show the
             Tresse and Maddox function multiplied by 0.18 which 
             approximately matches the USEL H$\alpha$ luminosity function 
             at the higher redshift.
  \label{lum-ha}}
  \end{centering}
  \end{figure}

We next constructed the H$\alpha$ luminosity functions of the USEL
sample following the procedures used in \citet{kakazu} but using the
present larger spectroscopic sample.  We used the narrow band magnitudes
to determine the line strength of the selection line; H$\alpha$ at low
redshift and \oiii$\lambda$5007 at high redshift. For the higher
redshift objects we computed the $H\beta$ flux from the
\oiii$\lambda$5007 flux and converted this to an $H\alpha$ flux using
the Case B ratio.  Because of the high observed frame equivalent widths
the primary fluxes are insensitive to the continuum determination.
However, they do depend on the filter response at the emission line
wavelength so we restricted ourselves to redshifts where the nominal
filter response is greater than 80\% of the peak value following the
procedure used in the equivalent width analysis.  This also has the
advantage of providing a uniform selection and we assume the window
function is flat over the defined redshift range.  Now the volume is
simply defined by the selected redshift range for all objects above the
minimum luminosity which we take as corresponding to an observed flux of
$1.5\times10^{-17}$ erg cm$^{-2}$ s$^{-1}$ in the $H\alpha$ line.  For
the high redshift sample selected in the \oiii$\lambda$5007 line the
true flux limit will generally be lower than this since the
\oiii$\lambda$5007 line is normally stronger than H$\alpha$ and we
restrict the sample to objects with $H\alpha$ line fluxes lying above
this threshold. The luminosity function is now obtained by dividing the
number of objects in each luminosity bin by the volume.  The
incompleteness corrected luminosity function is obtained from the sum of
the weights in each luminosity bin divided by the volume. Here the
weight of an object of a given magnitude correponds to the total number
of objects at that magnitude divided by the number of identified ojects
at that magnitude. Because of the high spectroscopic completeness the
incompleteness corrections are small except at the very lowest
luminosities. The 1 sigma errors shown are calculated from the
Poissonian errors based on the number of spectroscopically identified
objects in the bin.  The calculated H$\alpha$  luminosity function  is
shown for the $z=0.2-0.45$ range corresponding to the NB816 and NB912
H$\alpha$ selections in Figure~\ref{lum-ha}(a) and the corresponding
H$\alpha$ luminosity function at $z=0.6-0.9$ from the \oiii$\lambda$5007
selected samples in Figure~\ref{lum-ha}(b).

The USEL H$\alpha$ function at $z=0.3$ corresponds to about 4\% of the
total H$\alpha$ luminosity function at this redshift \citep{tresse98}
which is shown as the blue dashed line in Figure~\ref{lum-ha}(a).  This
is similar to the fraction of the total star formation rate that
\citet{kakazu} estimated was in the USELs. The USEL H$\alpha$ luminosity
function rises rapidly with $z=0.3$ paralleling the rapid rise in the
star formation rates over this redshift interval.


  \begin{figure}
  \vspace*{-0.2in}
  \begin{centering}
    \includegraphics[width=3.4in]{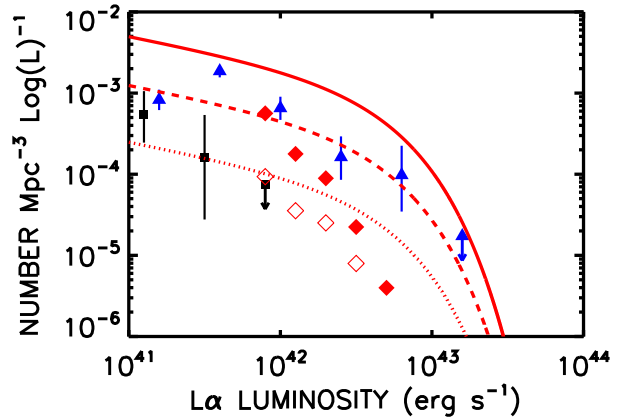}
    \caption{The Ly$\alpha$ luminosity functions computed from the 
             present samples assuming f(Ly$\alpha$)/f(H$\alpha$)$=5$
             is shown at $z=0.3$ (black squares) and at $z=0.7$ (blue
             triangles). Only objects with rest frame equivalent widths
             above 20\AA\ are included. The error bars are one sigma
             computed from the Poisson errors corresponding to the
             number of spectroscopically observed objects in the bin.
             We also show the one sigma upper limit (downward
             pointing arrow) at the highest luminosity.  The Ly$\alpha$ 
             luminosity function at the  $z=0.3$ redshift derived from 
             GALEX observations by  \citet{deh08} is shown with the open 
             red diamonds (actual obervations) and solid red diamonds 
             (incompleteness corrected).  It is clear that there are few 
             local high luminosity LAEs in an area of the present size 
             and there is no strong overlap between the present sample 
             and the GALEX sample. We also show the $z=3$ LAE luminosity 
             function of \citet{gronwall07} (solid red line) and the same 
             function with a downward scaling of 4 (dashed red line) and 
             20 (dotted red line) in the number density.
  \label{lf_comp}}
  \end{centering}
  \end{figure}

These H$\alpha$ luminosity functions also allow us to construct upper
limits on the $z=0-1$ Lyman alpha emitter luminosity functions for
comparson with the local LAE luminosity function from GALEX
\citep{deh08} and with the measurements of the LAE luminosity function
at high redshift ($z=2-7$).  As we discuss in more detail below a galaxy
with a high equivalent width in the Ly$\alpha$ line will also have a
high equivalent width in H$\alpha$, so the LAEs are a subset of the
present sample.  This allows us to construct a upper bound on the
complete LAE sample to a fixed Ly$\alpha$ equivalent width which can
then be compared to the local and high redshift population in detail.

As is well known, the short path length to optical scattering on neutral
hydrogen ensures that Lyman alpha photons follow a complex escape from a
galaxy.  While the low metallicity and extinction of the present USEL
sample are clearly positive indicators of a higher Ly$\alpha$ escape
fraction they are by no means the only factor. Internal structure,
kinematics, and the distribution of star formation sites may also play
key roles and may in fact be the more critical determinants. We take the
Case B ratio of $8-12$ for the Ly$\alpha$/H$\alpha$ flux, depending on
the electron density, to be an upper bound on the line ratio
\citep{ost85} (though we note that it is possible to have geometries in
which the scattering can enhance the Ly$\alpha$ line relative to the UV
continuum \citep{fink07}.) Given the roughly flat $f_\nu$ continua in
these galaxies (or $f_\lambda \sim \lambda^{-2}$), and the rest-frame
fully complete to a rest equivalent-width of 30\AA\ compared to the
20\AA\ equivalent width normally used to select the high-redshift LAE
population. The higher redshift $z\sim0.7$ sample will contain all LAEs
above 20\AA\ for the case B assumption. However, empirical estimates of
the Ly$\alpha$ escape fraction in the high redshift LAE population made
by comparing star formation rates estimated from the UV continuum with
those from the Ly$\alpha$ line suggest escape fractions of 30-50\%
\citep{gaw07,gaw08}.  If this lower ratio is applicable for the ratio of
the Ly$\alpha$/H$\alpha$ fluxes.  then all LAEs will be included in the
present samples, though objects close to the H$\alpha$ limiting
equivalent width will fall from the LAE sample. The LAE sample is
therefore a subset of the present sample and we can directly obtain an
upper bound on the LAE luminosity function from the USEL H$\alpha$
luminosity function.

The sample is small but it provides a first-cut estimate of the
evolution of the LAE luminosity function. We illustrate this in
Figure~15 where we show the $z=0.3$ and the $z=0.7$ LAE luminosity
functions that would be derived from our H$\alpha$ sample under the
assumption Ly$\alpha$/H$\alpha$$=5$ for objects with rest frame
Ly$\alpha$ equivalent widths above 20\AA.  The $z=0.3$ LAE luminosity
function cam be compared with the direct determinations of the LAE
luminosity function at this redshift from the GALEX observations of
\citet{deh08} which are shown with the open red diamonds.  The GALEX
data only contain objects with strong UV continuum detections and these
correspond to the highest luminosity emitters.  In this sense they
parallel the high redshift Lyman break galaxies with strong Lyman alpha
emission rather than the line selected LAE samples. \citeauthor{deh08}
estimated that there should be a large incompleteness correction to
allow for this selection effect and the solid red diamonds show their
incompleteness corrected luminosity function. However, the present data,
even under the case B assumption, suggest that the incompleteness
corrections at the low luminosity end are smaller than the \citet{deh08}
estimate.  (We recall that the present data provide an absolute upper
limit for the case B assumption.) The present $z=0.3$ sample has too
small a volume to probe the high luminosity end.

At high redshifts the LAE luminosity function is surprisingly invariant
in the range $z=2.5-6$\ but there are signs it begins to drop at both
higher and lower redshift. The present data show that it must have
dropped substantially at redshift $z=0.7$ and even further at $z=0.3$.
In Figure~\ref{lf_comp} we show the most recent determination of the
$z=3$ LAE luminosity function from \citet{gronwall07}.  In order to
match the present data the LAE luminosity function would have to drop by
about a factor of four between $z=3$ and $z=0.7$ (dashed red line) and
by a factor of very roughly twenty between $z=3$ and $z=0.3$ (dotted red
line). Even under the case B assumption these drops would be two and ten
respectively. The drop in the LAE luminosity fuction parallels the drop
in the star formation over this redshift interval and there is even a
suggestion of donwsizing in the shape of the LAE LF at the lowest
redshift which is deficient in high luminosity objects relative to the
low luminosity end.

\section{Summary}
\label{secsum}

We have described the results of deep spectroscopic observations of a
narrow band selected sample of extreme emission line objects. The
results show that such objects are common in the $z=0-1$ redshift
interval and that a very large fraction of the strong emitters are
detected in the \oiii$\lambda$4363 line where oxygen abundances can be
measured using the direct method.  The abundances primarily lie in the
12+log(O/H) range of 7$-$8 characteristic of XMPGs. We have determined
the metal-luminosity relation for this class of object finding it lies
about 0.6 dex below a magnitude selected sample in the same redshift
interval.  We give an emprical relation between R23 and 12 + log(O/H)
which differs from local estimates. We also show that low metallicity
objects can be picked out by the weakness of
\oii$\lambda$3727/\oiii$\lambda$5007 in the spectra. The two lowest
metallicity galaxies in the sample have 12 + log(O/H) = $6.97 \pm 0.17$
and $7.25 \pm 0.03$,  making them among the lowest metallicity galaxies
known, but we expect that as the sample size is increased yet lower
metallicity galaxies may be found and that we may hope to be able to
determine if there is a floor in the galaxy metallicity at these
redshifts.

\acknowledgements
We are indebted to the staff of the Subaru and Keck
observatories for their excellent assistance with the
observations. We would also like to thank Rolf Kudritzki
for a thoughtful reading of the first draft of this paper.
We gratefully acknowledge support from NSF grants AST-0607850 (E.~M.~H.)
AST-0709356 (L.~L.~C.), and AST-0708793 (A.~J.~B.),
the University of Wisconsin Research Committee with funds
granted by the Wisconsin Alumni Research Foundation,
and the David and Lucile Packard Foundation (A.~J.~B.).

{\it Facilities:} \facility{Keck:II (DEIMOS)}, \facility{Subaru (SuprimeCam)}

%
%

\end{document}